\documentclass[aps,prb,twocolumn,superscriptaddress,showpacs,amsmath,amssymb]{revtex4}

\usepackage{graphicx}

\begin{document}

\title{Discrete antiferromagnetic spin-wave excitations in the giant ferric wheel Fe$_\text{18}$}

\author{J. Ummethum}
\affiliation{Fakult\"at f\"ur Physik, Universit\"at Bielefeld, Postfach 100131, D-33501 Bielefeld, Germany}

\author{J. Nehrkorn}
\affiliation{Physikalisches Institut, Universit\"at Freiburg, D-79104 Freiburg, Germany}

\author{S. Mukherjee}
\affiliation{Department of Chemistry, University of Florida, Gainesville, Florida 32611, USA}

\author{N. B. Ivanov}
\affiliation{Fakult\"at f\"ur Physik, Universit\"at Bielefeld, Postfach 100131, D-33501 Bielefeld, Germany}
\affiliation{Institute of Solid State Physics, Bulgarian Academy of Sciences, Tzarigradsko chaussee 72, 1784 Sofia, Bulgaria}

\author{S. Stuiber}
\affiliation{Physikalisches Institut, Universit\"at Freiburg, D-79104 Freiburg, Germany}

\author{Th. Str\"assle}
\affiliation{Laboratory for Neutron Scattering, Paul Scherrer Institut, CH-5232 Villigen PSI, Switzerland}

\author{P. L. W. Tregenna-Pigott}
\affiliation{Laboratory for Neutron Scattering, Paul Scherrer Institut, CH-5232 Villigen PSI, Switzerland}

\author{H. Mutka}
\affiliation{Institut Laue-Langevin, 6 rue Jules Horowitz, BP 156, F-38042 Grenoble Cedex 9, France}

\author{G. Christou}
\affiliation{Department of Chemistry, University of Florida, Gainesville, Florida 32611, USA}

\author{O. Waldmann}
\affiliation{Physikalisches Institut, Universit\"at Freiburg, D-79104 Freiburg, Germany}

\author{J. Schnack}
\affiliation{Fakult\"at f\"ur Physik, Universit\"at Bielefeld, Postfach 100131, D-33501 Bielefeld, Germany}

\date{\today}

\begin{abstract}
The low-temperature elementary spin excitations in the AFM molecular wheel Fe$_{18}$ were studied experimentally by
inelastic neutron scattering and theoretically by modern numerical methods, such as dynamical density matrix
renormalization group or quantum Monte Carlo techniques, and analytical spin-wave theory calculations. Fe$_{18}$
involves eighteen spin-5/2 Fe$^\text{III}$ ions with a Hilbert space dimension of $\sim10^{14}$, constituting a
physical system that is situated in a region between microscopic and macroscopic. The combined experimental and
theoretical approach allowed us to characterize and discuss the magnetic properties of Fe$_{18}$ in great detail. It is
demonstrated that physical concepts such as the rotational-band or $L \& E$-band concepts developed for smaller rings
are still applicable. In particular, the higher-lying low-temperature elementary spin excitations in Fe$_{18}$ or AFM
wheels in general are of discrete antiferromagnetic spin-wave character.
\end{abstract}


\maketitle

\section{\label{sec:intro}Introduction }

\begin{figure}
\includegraphics{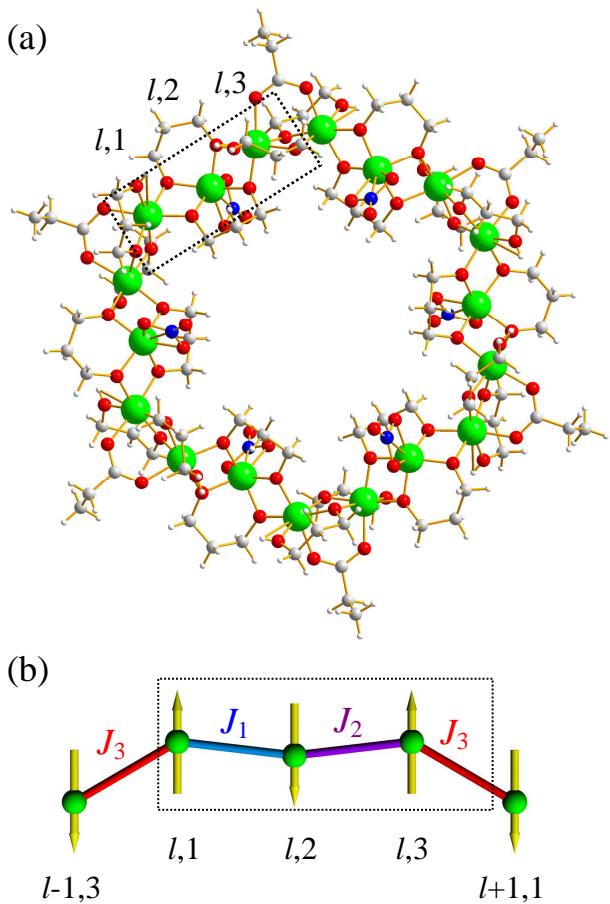}
\caption{\label{fig:molecule}
(Color online) (a) Ball-and-stick representation of the molecular structure of the Fe$_{18}$ molecule (Fe: green, O:
red, N: blue, C: gray, and H: white). The dashed box indicates the unit cell used for magnetic modeling, see next panel. (b) Unit
cell (dashed box) and labeling of the spin sites (text at bottom) used for the theoretical calculations as suggested by
the symmetry of the molecule. The exchange constants $J_1$, $J_2$, and $J_3$ associated to different bonds are
indicated.}
\end{figure}

Ring-like arrangements of a dozen or so of magnetic spins experiencing nearest-neighbor antiferromagnetic (AFM)
exchange interactions, as realized experimentally for instance by the AFM molecular wheels, have attracted significant
attention in the past
decade.\cite{mag_af_ring,SBUH:AngewChemIntEd1997,Wetal:AngewChemIntEd1997,Chi98,ACCFG:InorgChimActa2000,BSS:JMMM00:B,vSetal:ChemEurJ2002,BHS:PRB03,Wal05-gridreview,Wal03-cr8,MFK:PRL06,HNN:JACS09}
The molecular ferric wheel [Fe$_\text{18}$(pdH)$_\text{12}$(O$_\text{2}$CEt)$_\text{6}$(NO$_\text{3}$)$_\text{6}$], or
Fe$_{18}$ in short, is the largest magnetic molecular wheel synthesized to date.\cite{fe18synthese} The molecule
contains $N = 18$ Fe$^\text{III}$ ions with spin $s = 5/2$, arranged in a ring-like fashion as shown in
Fig.~\ref{fig:molecule}(a), and exhibits crystallographic $C_6$ symmetry. Its large yet finite size makes it an ideal
candidate to explore the region between microscopic and macroscopic physics (the system is mesoscopic). The magnetism
in the Fe$_{18}$ wheel was studied before using high-field magnetic torque measurements, and the low-lying energy
spectrum up to 2~meV was investigated by inelastic neutron scattering (INS).\cite{Wal09-fe18} The experimental data
demonstrated the dynamics of the N\'eel vector, and the magnetic torque provided direct evidence for quantum
oscillations in the N\'eel vector tunneling gap due to quantum phase interference.\cite{Wal09-fe18}

The experimental observations have been well described in terms of an effective two-sublattice Hamiltonian, which had
been demonstrated before for smaller wheels with up to 10 spin sites to approximate well the low-energy part of the
true spectrum.\cite{ScL:PRB00,Wal02-spindyn} This Hamiltonian is in fact related to the more general concept of the $L
\& E$-band picture, which describes the elementary excitations as a set of rotational (parabolic) energy bands, and
which was shown to apply to a variety of AFM spin clusters with bipartite or tripartite sublattice
structure.\cite{ScL:PRB00,SLM:EPL01,Wal02-spindyn,Dre10-csfe8ins3,Wal07-swtfe30,Wal03-cr8} The effective Hamiltonian
allowed a description of the experiments because it operates in a Hilbert space of dimension 2116 (for Fe$_{18}$),
which can easily be handled on a personal computer using standard numerical diagonalization techniques. However, direct
confirmation of its applicability to wheels as large as Fe$_{18}$ is lacking and important magnetic parameters such as
magnetic anisotropy could not be determined reliably. Furthermore, higher-lying spin excitations, which are expected in
the $L \& E$-band concept, were not observed.

The numerically exact evaluation of all energy eigenvalues and eigenfunctions of a giant molecule such as Fe$_{18}$
poses a great challenge for theory since the size of the related Hilbert space grows as $(2s+1)^N$ and for Fe$_{18}$
assumes a value of $101,559,956,668,416 \approx 10^{14}$. This dimension is much too big for a matrix diagonalization.
Even a decomposition of the Hamiltonian matrix according to the available symmetries, as successfully done for the
smaller AFM wheels such as CsFe$_{8}$ or Fe$_{10}$,\cite{DGP:IC93,Wal00-sym,ScS:IRPC10,Pil05} is not efficient enough
to ease the problem. Fortunately, the magnetic molecule Fe$_{18}$ is non-frustrated, which permits the application of
Quantum Monte Carlo (QMC) methods,\cite{SaK:PRB91,San:PRB99,ALPS:JMMM07} and furthermore is quasi one-dimensional,
which makes it ideal for (Dynamical) Density Matrix Renormalization Group (DDMRG \& DMRG)
calculations.\cite{Whi:PRL92,Whi:PRB93,Sch:RMP05,CorrVecRamasesha,CorrVecKuehnerWhite,DDMRG} The first method allows
the evaluation of thermodynamic observables such as the magnetic susceptibility whereas the second delivers transition
rates between low-lying energy levels that can be related to the INS spectrum.

In this work a comprehensive study of the higher-lying excitations in Fe$_{18}$, which in the language of the $L \&
E$-band concept correspond to the $E$ band or discrete spin-wave excitations, is reported. Experimentally the
excitation spectrum was determined by high-energy INS measurements extending the energy range to 13.5~meV, and the
temperature-dependent magnetic susceptibility, which probes the full energy spectrum. Theoretically, the QMC and DDMRG
techniques were used to reproduce the experimental magnetic susceptibility and INS data with excellent accuracy. Because of the size and the structure of Fe$_{18}$ the DDMRG calculations are time-consuming, which prevents a systematic scanning of
the magnetic parameter space or least-squares fit approaches. This problem was circumvented by resorting to spin-wave
theory as an intermediate step. This allowed us to refine the microscopic spin Hamiltonian for Fe$_{18}$ and deduce
accurate microscopic magnetic parameters. A key result is that the spin-wave like character of the higher lying
excitations is preserved although the refined Hamiltonian is less symmetric compared to the assumptions in
Ref.~\onlinecite{Wal09-fe18}. Furthermore, the validity of the $L \& E$-band concept is confirmed for Fe$_{18}$.

The article is organized as follows: It begins with a discussion of the experimental results for the susceptibility and
the INS cross section in Sec.~\ref{sec:exp}. This is followed by a theoretical analysis in Sec.~\ref{sec:analysis} and
a discussion in Sec.~\ref{sec:discussion}. After our final conclusions further details of the employed methods are
given in an appendix.

\section{\label{sec:exp} Experimental Results}

\begin{figure}
\includegraphics{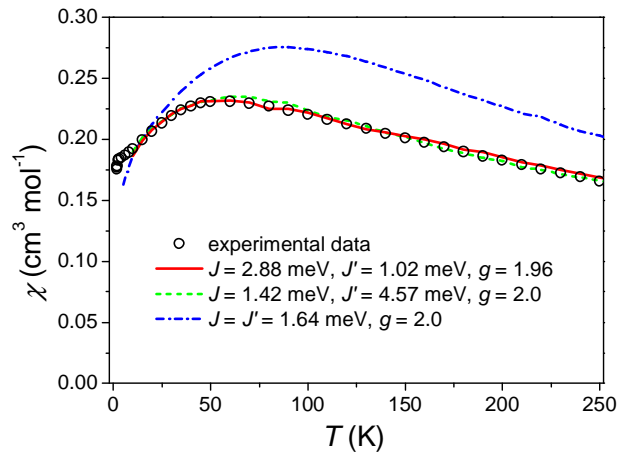}
\caption{\label{fig:xi} (Color online) Experimental (open circles) and simulated (lines) magnetic susceptibility as
function of temperature. The three simulated curves were obtained from the microscopic model
Eq.~(\ref{eq:heisenberg}) using QMC. The parameters in the simulations are as indicated (with $J_1=J_2=J$ and $J_3=J'$). }
\end{figure}

The experimental methods used for sample preparation, and magnetic susceptibility and INS measurements are described in
Appendix~\ref{sec:methodsexp}. Figure~\ref{fig:xi} shows the magnetic susceptibility $\chi$ as function of temperature
measured on a poly-crystalline sample. At 250~K, a $\chi$ value of 0.17~cm$^3$/mol is observed, which increases with
decreasing temperature, reaches a maximum of 0.23~cm$^{3}$/mol at a temperature of ca. 50~K, and decreases further with
decreasing temperature. At the lowest temperature an abrupt decrease is observed. Such a behavior is typical for AFM
molecular wheels,\cite{mag_af_ring} and the data are consistent with earlier measurements on a micro-crystalline
sample.\cite{fe18synthese} The ground state of an even-membered AFM wheel is a total spin singlet, $S=0$, and the
susceptibility is hence expected to approach zero at zero temperature, which is not observed here because in large
wheels the drop to $\chi=0$ occurs only at the lowest temperatures (the gap to the first excited triplet is roughly
given by $\sim 4J/N$ and is small in Fe$_{18}$) and the susceptibility in this temperature regime is strongly affected
by the presence of magnetic anisotropy (which is significant in Fe$_{18}$).\cite{Wal99-fe6} The solid curves are
theoretical results which are discussed below.

\begin{figure}
\includegraphics{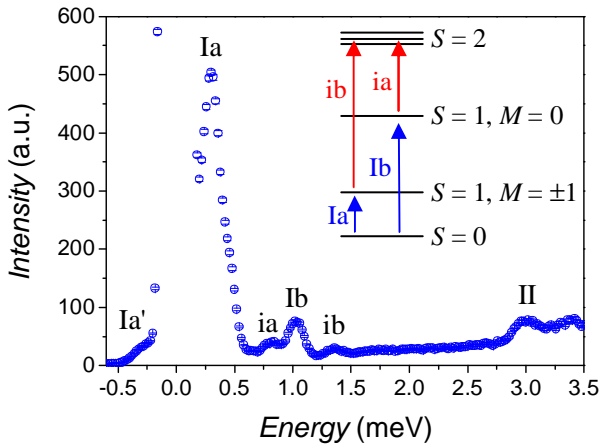}
\caption{\label{fig:INS_4p2A}
Experimental INS spectra recorded on IN5 with incoming neutron wavelength $\lambda$ = 4.2~{\AA} at 1.9~K. Positive
energy transfer corresponds to neutron energy loss. The labels indicate observed transitions. The inset sketches the
assignment of the low-energy excitations as inferred previously.\cite{Wal09-fe18} }
\end{figure}

The INS spectrum recorded at an incoming wavelength of $\lambda$ = 4.2~{\AA} and a temperature of 1.9~K on the
spectrometer IN5 is shown in Fig.~\ref{fig:INS_4p2A}. At low energies a strong feature at 0.3~meV (peak Ia) together
with its corresponding anti-Stokes feature (peak Ia') is observed. At around 1~meV a group of three features appears,
peak ia at 0.8~meV, peak Ib at 1~meV, and peak ib at 1.36~meV. These transitions were already observed in the previous
low-energy INS experiment\cite{Wal09-fe18} and interpreted as follows: Peaks Ia and Ib correspond to cold magnetic
transitions from the $S = 0$ ground state to the first excited $S =1$ multiplet, which is zero-field split by magnetic
anisotropy into its components $M = \pm 1$ and $M = 0$, and peaks ia and ib were identified as hot magnetic transitions
from this first excited $S = 1$ multiplet to the next-higher lying $S = 2$ multiplet (see inset to
Fig.~\ref{fig:INS_4p2A}). In addition to these transitions, a further peak II at 3.0~meV is observed.

\begin{figure}
\includegraphics{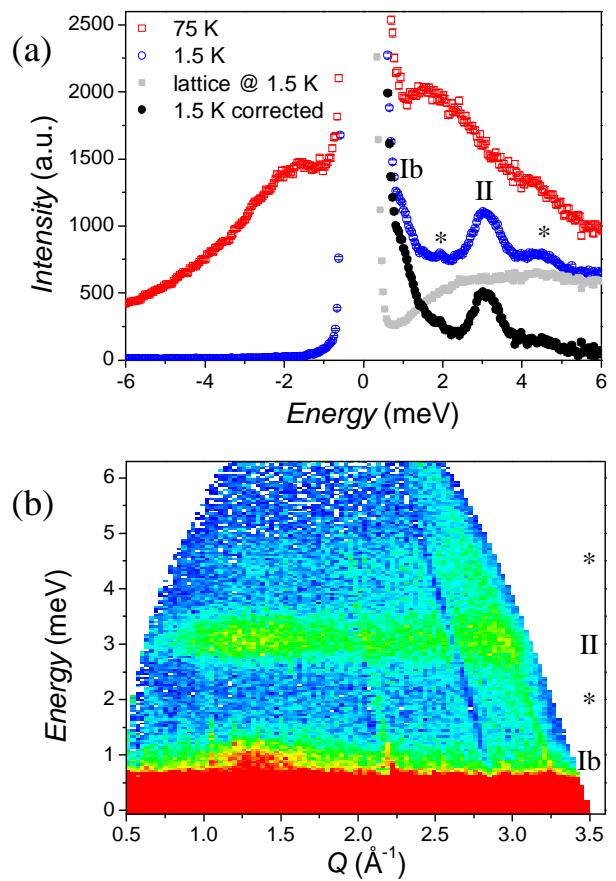}
\caption{\label{fig:INS_3p2A}(Color online)
(a) Experimental INS spectra recorded on FOCUS with incoming neutron wavelength $\lambda$ = 3.2~{\AA} at 1.5 (blue open
circles) and 75~K (red open squares). The 75~K data were Bose-scaled to yield the estimated lattice scattering at 1.5~K
(gray solid squares) and were subtracted from the 1.5~K data yielding the Bose-corrected data (black solid circles).
(b) $S(Q,\omega)$ plot of the (not Bose-corrected) 1.5~K data shown in panel (a). Intensity is color-coded from blue (low intensity) to red
(high intensity). The labels indicate observed transitions, and the asterisks spurion features as discussed in the
text. Positive energy transfer corresponds to neutron energy loss.}
\end{figure}

Figure~\ref{fig:INS_3p2A}(a) presents the INS spectra recorded with an incoming wavelength $\lambda$ = 3.2~{\AA} on the
spectrometer FOCUS. In the 1.5~K data a prominent feature at 3~meV is observed on the neutron-energy loss side, which
obviously corresponds to peak II found before in the IN5 data (Fig.~\ref{fig:INS_4p2A}). Also, a shoulder near the
elastic line at ca. 1~meV is observed, which obviously corresponds to peak Ib (peak ib is not detected here because of
its weak intensity and the lower experimental resolution in the $\lambda$ = 3.2~{\AA} experiment). Two further features
are observed, a very weak feature at ca. 2~meV and a weak broad feature at ca. 4.5~meV. The spectrum recorded at 75~K
shows, apart from a strongly increased background due to the excited lattice, no clear features. It is therefore
reasonable to assume that at this temperature predominantly the lattice excitations are observed, and the magnetic
scattering intensity is distributed over all energies. The lattice contribution (on the neutron-energy loss side) at
low temperatures may thus be estimated by scaling the 75~K data with the Bose factor $[1-\exp(-E/k_B T)]^{-1}$, which
determines the temperature dependence of phononic scattering. The estimated lattice contribution is then subtracted
from the low-temperature INS data, a procedure we call Bose correction. This approach was used with considerable
success in the
past.\cite{Och08-cr6cr7,Dre10-csfe8ins3,Stu11-mn10} In the Bose-corrected 1.5~K data, shown also in
Fig.~\ref{fig:INS_3p2A}(a), peak II remains strong, providing a strong hint that it is of magnetic origin.

Also the dependence of the INS intensity on momentum transfer $Q$ could be studied, which often allows for an
unambiguous conclusion as regards the origin of INS features. The $S(Q,\omega)$ plot of the 1.5~K data is shown in
Fig.~\ref{fig:INS_3p2A}(b). Peaks Ib and II are clearly identified, which is impressive considering that a
non-deuterated molecular powder sample was measured at high energies. Magnetic and phononic excitations may be clearly
differentiated by their $Q$ dependence, since for the former the intensity is either strongest at low $Q$ values or
typically is maximal at around 1.2~{\AA}$^{-1}$,\cite{Wal03-insqdependence} while for the latter an increase of the
intensity with $Q^2$ is expected.\cite{ts_insformel} Clearly, the intensity of both peaks Ib and II is negligible at
low $Q$ and passes through a maximum between 1.1 and 1.5~{\AA}$^{-1}$. At higher $Q$ values the observed intensity is
almost constant. Such a $Q$ dependence is characteristic for the magnetic excitations in AFM molecular
wheels.\cite{Wal99-fe6,Car03-cr8,Wal03-cr8,Wal05-csfe8ins1,San05-fe10nvt,Wal06-csfe8ins2,Dre10-csfe8ins3} Hence, on the
basis of the temperature dependence, the Bose correction, and the $Q$ dependence peaks Ib and II are clearly magnetic.
In contrast, the features at 2~meV and 4.5~meV do exhibit their strongest intensity at large $Q$ values, and are hence
safely assigned to spurious and/or lattice contributions, which for the 2~meV feature is also confirmed by its absence
in the $\lambda$ = 4.2~{\AA} data, Fig.~\ref{fig:INS_4p2A}.

\begin{figure}
\includegraphics{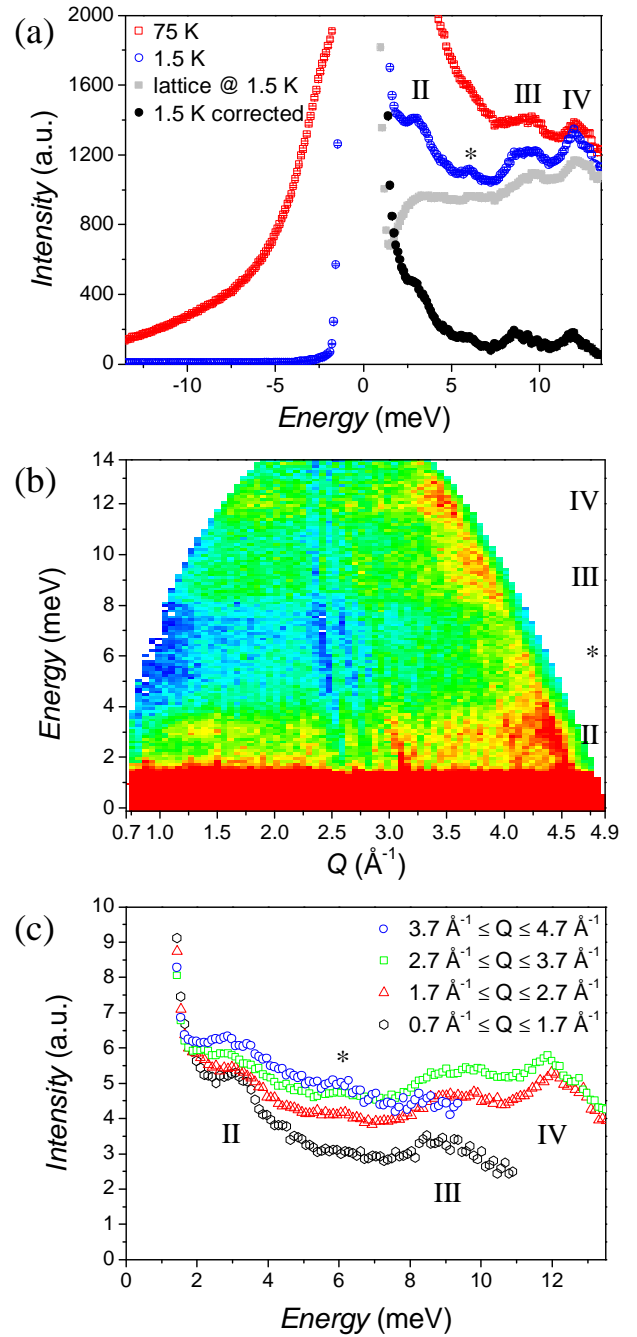}
\caption{\label{fig:INS_2p26A}(Color online) (a) Experimental INS spectra recorded on FOCUS with incoming neutron
wavelength $\lambda$ = 2.26~{\AA} at 1.5 (blue open circles) and 75~K (red open squares). The 75~K data were
Bose-scaled to yield the estimated lattice scattering at 1.5~K (gray solid squares) and were subtracted from the 1.5~K
data yielding the Bose-corrected data (black solid circles). (b) $S(Q,\omega)$ plot of the (not Bose-corrected) 1.5~K data shown in panel
(a). Intensity is color-coded from blue (low intensity) to red (high intensity). (c) $Q$-sliced 1.5~K data, with the
$Q$ slices as indicated. The labels indicate observed transitions and the asterisk a spurion feature as discussed in
the text. Positive energy transfer corresponds to neutron energy loss.}
\end{figure}

The INS spectra recorded on FOCUS with incident wavelength $\lambda$ = 2.26~{\AA} at 1.5 and 75~K are displayed in
Fig.~\ref{fig:INS_2p26A}(a). In the 1.5~K spectrum peak II at ca. 3~meV is again observed. Furthermore, prominent
features are also observed at around 8.5~meV (peak III) and 12~meV (peak IV), and at ca. 6~meV a weak feature is found.
Peak III is significantly broader than the experimental resolution, and appears to consist of two features. The 75~K
spectrum exhibits an increased intensity reflecting the strong phononic scattering intensity, and no detailed features
are observed at the lower energies. However, two features are present at energies roughly corresponding to those of
peaks III and IV. A Bose correction as discussed before yielded the Bose-corrected 1.5~K data shown in
Fig.~\ref{fig:INS_2p26A}(a). Both peaks III and IV are pronounced in the Bose-corrected data, which strongly hints
towards a magnetic origin of this scattering intensity. The right shoulder at ca. 9~meV on the high-energy side of
feature III in the (original) 1.5~K data is significantly reduced by the Bose-correction, which suggests a lattice
origin of this scattering intensity. The $S(Q,\omega)$ plot of the 1.5~K spectrum is given in
Fig.~\ref{fig:INS_2p26A}(b). The peaks II, III, and IV are clearly present and exhibit strong scattering intensity at
low $Q$ values, with indications of a maximum below 1.5~{\AA}$^{-1}$, which unambiguously demonstrates their magnetic
origin. At all energies a significant phonon contribution is observed at the largest $Q$ values. To gain further
insight into the nature of the weak feature at ca. 6~meV, the INS data were analyzed via $Q$ slices, as shown in
Fig.~\ref{fig:INS_2p26A}(c). The 6~meV feature is the better detected in the $Q$ slices the higher the $Q$ values are,
and appears to be essentially absent in the $Q$ slice with the lowest $Q$ values. It is hence assigned to a spurion or
lattice feature. In contrast, the peaks II, III, and IV are more pronounced in the lower $Q$ slices, again
demonstrating their magnetic origin.

To summarize the INS findings, besides the four low-energy features Ia, Ib, ia, and ib, which were already observed
before in previous INS experiments,\cite{Wal09-fe18} three further cold magnetic transitions II, III, and IV were
observed in the high-energy regime up to 14~meV. An analysis of these transitions using Gaussian fits with sloped
background yielded the energy positions as peak II: 3.0(1)~meV, peak III: 8.5(2)~meV, peak IV: 12.0(2)~meV.

\section{\label{sec:analysis} Analysis}

\subsection{\label{sec:model} Spin model for Fe$_\text{18}$}

The generic Hamiltonian for the description of the magnetic properties of a molecular wheel of $N=18$ Fe$^\text{III}$
spins is a Heisenberg Hamiltonian with nearest-neighbor interactions plus a term reflecting the single-ion anisotropy
of the Fe$^\text{III}$ ions.\cite{Wal09-fe18} In principle, this would lead to at least 36 unknown parameters. However,
Fe$_\text{18}$ possesses a crystallographic $C_6$ symmetry axis perpendicular to the wheel plane, which leads to a
repeating unit of three Fe$^\text{III}$ ions, as indicated in Fig.~\ref{fig:molecule}(b). Therefore, the Heisenberg
part of the Hamiltonian can be formulated with at most three different exchange constants:
%
\begin{equation}
\label{eq:heisenberg}
\widehat{H}_{H}=\sum_{l=1}^{L=6} J_1\mathbf{\widehat{S}}_{l,1}
\cdot \mathbf{\widehat{S}}_{l,2} + J_2\mathbf{\widehat{S}}_{l,2}
\cdot \mathbf{\widehat{S}}_{l,3} +
J_3\mathbf{\widehat{S}}_{l,3}\cdot \mathbf{\widehat{S}}_{l+1,1}\,,
\end{equation}
%
where $L = N/3$ is the number of unit cells and $l$ enumerates the repeating units and has to be understood modulo $L$
[for the enumeration of the individual spin sites see also Fig.~\ref{fig:molecule}(b)].

The magnetic susceptibility, Fig.~\ref{fig:xi}, is compatible with a non-magnetic, total spin $S=0$ ground state. In
addition, the high temperature behavior of $\chi$ requires that the sum of all exchange interactions is
antiferromagnetic.\cite{SSL:PRB01} Considering the susceptibility function further, one can even conclude that all
couplings along the ring have to be antiferromagnetic because if any of the exchange interactions were ferromagnetic,
either the ground state would possess a total spin of $S=15$ or the susceptibility would rise to much larger values at
low temperatures than observed. Hence, we can safely restrict our analysis to AFM exchange interactions, $J_1,J_2,J_3 >
0$. A closer look at the structure of the molecule suggests the simpler model where $J_2 = J_3 = J$ and $J_1 = J'$,
since the exchange bridges connecting the centers $\widehat{S}_{l,1}$ to $\widehat{S}_{l,2}$ and $\widehat{S}_{l,2}$ to
$\widehat{S}_{l,3}$ are chemically identical and structurally very similar, while the bridge connecting centers
$\widehat{S}_{l,3}$ to $\widehat{S}_{l+1,1}$ is chemically very different (see Fig.~\ref{fig:molecule}). In the
following we will denote the uniform ring with $J_1=J_2=J_3=J$ as a one-$J$ model, the situation with $J_2 = J_3 = J$
and $J_1 = J'$ as a two-$J$ model, and the general case of three different exchange constants as a three-$J$ model.

The interaction with an applied magnetic field  $\mathbf{B}$ is described by the Zeeman Hamiltonian
%
\begin{equation}
\widehat{H}_Z= g \mu_B \sum_{j=1}^{N=18}\mathbf{\widehat{S}}_j \cdot \mathbf{B}\,,
\label{eq:zeeman}
\end{equation}
%
where the $g$ factor is close to 2 for Fe$^\text{III}$ ions, $j$ enumerates the individual spin sites, and $\mu_B$ is
the Bohr magneton. Previous results suggest that the single-ion anisotropy of the Fe$^\text{III}$ ions, which is
modeled by the term
%
\begin{equation} \label{eq:fe18-anis-term}
	\widehat{H}_{D}=D\sum_{j=1}^{N=18}(\widehat{S}_{j}^{z})^{2}\,,
\end{equation}
is relatively small in magnitude yet strongly affects the spin dynamics at very low energies.\cite{Wal09-fe18} However,
it is expected to have negligible effect on the excitations at higher energies.\cite{Dre10-csfe8ins3} Therefore this
term is not included in our analysis, but this assumption is carefully checked and confirmed in Sec.~\ref{sec:expana}.

In order to determine the exchange parameters we approach the relevant energy spectrum in three steps: In a first step
we analyze the spin-wave excitations and compare them with the observed INS excitations in order to narrow down the range
of possible parameter values and to qualitatively understand the character of the excitations. This approach is
motivated by the $L \& E$-band picture, which connects the higher-energy low-temperature excitations to discrete
spin-waves, and spin-wave theory was indeed able to reproduce the excitations in the AFM wheel CsFe$_{8}$ with
semi-quantitative accuracy.\cite{Dre10-csfe8ins3} In a second step we perform large-scale DDMRG calculations for quite
a number of parameter sets that yield model parameters of high accuracy. Finally, these are compared with QMC
calculations of the magnetic susceptibility.

\subsection{\label{sec:swana} Spin-wave calculations}

The one-magnon excitations of Hamiltonian Eq.~(\ref{eq:heisenberg}) can be obtained in the framework of standard
spin-wave theory (SWT).\cite{IvS:LNP04} Some modification of the theory becomes inevitable, however, when studying
physical quantities such as the dynamical correlation function in Eq.~(\ref{eq:dynam_corr_func}), which exhibits
divergencies caused by the Goldstone modes. The gapped structure of these modes in finite Heisenberg clusters and AFM
wheels in particular can be handled within the framework of SWT by introducing chemical potentials for the spin sites,
yielding the so-called modified SWTs.\cite{And52-swt,oguchi,sw1,HT:PRB1989,sw3,imswt,SBMFT1,SBMFT2,GSG:PRB2006} We will
come back to this point in Sec.~\ref{sec:discussion}. Here we calculate the one-magnon spectrum of
Eq.~(\ref{eq:heisenberg}) in a linear SWT approximation; for details we refer to Appendix~\ref{sec:AppSW}.

\begin{figure}
\centering
\includegraphics{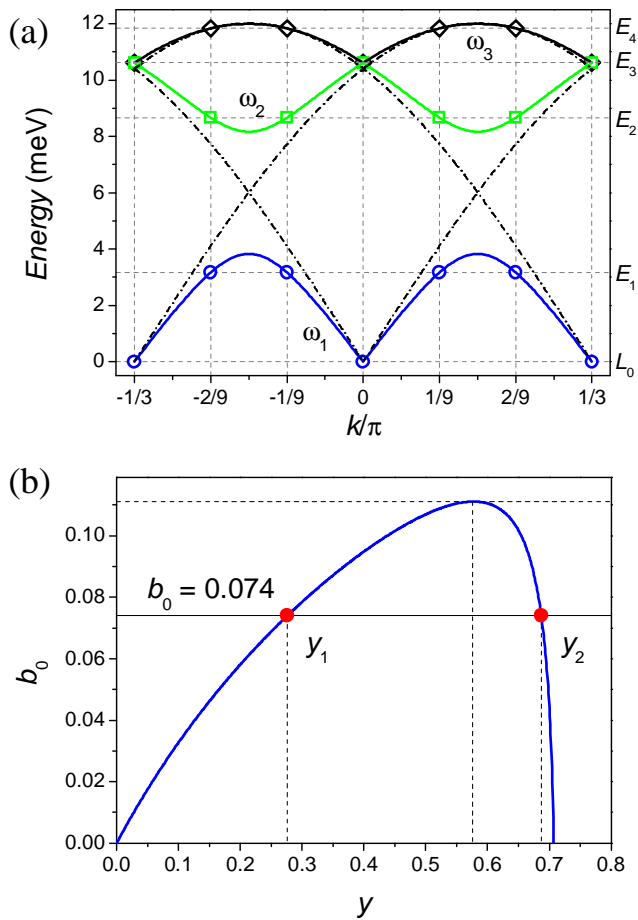}
\caption{\label{fig:sp}(Color online)
(a) Spin-wave excitation spectrum of Hamiltonian Eq.~(\ref{eq:heisenberg}) with $L=6$ (symbols) and $L=300$
(solid curves) as obtained from SWT for the parameter $b_0=0.074$. The energy levels $E_1$ to $E_4$ are
defined by Eq.~(\ref{ei}). The dashed-dotted curves display the
spectrum for the one-$J$ model.
(b) Dependence of Eq.~(\ref{b0}) for $J_1 = J_2 = J$ and $J_3 = J'$ on the scaling variable
$y$. The two solutions $y_1=0.276$ and $y_2=0.687$ for $b_0=0.074$ corresponding to $J'/J \approx 3$ and
$J'/J \approx 0.3$, respectively, are indicated. Both solutions produce identical
one-magnon spectra.}
\end{figure}

The spin-wave excitation energies of the Fe$_{18}$ system are displayed in Fig.~\ref{fig:sp}(a) as a function of a
shift quantum number $k$, which arises from Fourier transforming the periodic ring of $L$ unit cells (physically it
would correspond to a wavevector only in the infinite chain). For comparison, as guides to the eye, the dispersions of
the spin-wave branches of a large ($L=300$) ring are also shown. Three branches $\omega_{1}(k)$, $\omega_{2}(k)$,
$\omega_{3}(k)$ are observed as expected from the three centers in the unit cell. For Fe$_{18}$ the $k$ values are
restricted to the discrete values $k = 0, \pm \frac{1}{9} \pi, \pm \frac{2}{9} \pi, \frac{1}{3} \pi$, and a discrete
excitation spectrum with 17 energy states is obtained [$\omega_{1}(0)$ is the ground state]. The excitations fall into
four energy levels $E_1$, $E_2$, $E_3$, and $E_4$ (each is four-fold degenerate because of the symmetries in the SWT
approximation). The level $L_0$ is formed by the Goldstone mode $\omega_{1}(k=\pi/3)$, which has zero energy in the
standard SWTs. It acquires a gap however when modified SWTs are used, which in the quantum spectrum relates to the gap
between the $S=0$ ground state and first excited $S=1$ multiplet (singlet-triplet gap). Within the $L \& E$-band
concept the energies in the levels $E_1$ to $E_4$ correspond to the $E$ band, and the ground state and $k=\pi/3$ mode
to the $L$ band.\cite{Wal02-spindyn} The dispersions for the uniform ring or one-$J$ model are also displayed in
Fig.~\ref{fig:sp}(a) for comparison ($J$ was chosen here such that the maximal energy coincides).

The SWT calculations allow some useful insight. A modulation of the exchange constants away from uniform opens a gap
between the branches $\omega_{1}(k)$ and $\omega_{2}(k)$, which in relative terms shifts $E_1$ down and $E_2$ up. As
shown in Appendix~\ref{sec:AppSW}, SWT implies three relations between the reduced energies $E_1/E_3$, $E_2/E_3$ and
$E_4/E_3$, which depend only on the dimensionless parameter $b_0$,
%
\begin{equation}
\label{b0}
b_0 = \frac{3 J_1^2 J_2^2 J_3^2}{(J_1 J_2 + J_2 J_3 + J_3 J_1)^3}.
\end{equation}
%
Detailed inspection shows that the difference between the energy levels $E_3$ and $E_4$ does of course vary with
varying exchange constants, but always remains relatively small, and it turns out that it is always significantly too
small to account for the experimentally observed energy difference of peaks III and IV. Finally, the relation $E_1 +
E_2 = E_4$ follows from the first two equations in Eq.~(\ref{viete}). The conclusion from all this is that even in the
most general case of three different exchange couplings, the possible variation in the relative energies of the levels
is restricted. In particular, the three-$J$ model does not introduce significantly more flexibility in adjusting the
energies as compared to the two-$J$ model, despite the additional free parameter. Intuitively this is reasonable, since
the main characteristics which govern the dispersions in Fig.~\ref{fig:sp}(a) is the maximal energy (or $E_4$) and the
size of the gap between the energies $E_1$ and $E_2$. This strongly suggests that the possible small differences
between the exchange constants $J_1$ and $J_2$ cannot be resolved in the present experiments, and that therefore the
two-$J$ model is appropriate for Fe$_{18}$.

Comparing the spin-wave spectrum with the experimental INS peaks, it is obvious that the level $E_1$ corresponds to
peak II (peaks Ia and Ib are related to the singlet-triplet gap or level $L_0$, which in the present SWT calculation is
obtained at zero energy). However, at higher energies SWT predicts three further levels while only two INS peaks III
and IV are observed. This results in several possibilities for the assignment, which could all be ruled out because of
discrepancies with experiment except one, namely the assignment that $E_2$ corresponds to peak III and $E_4$ to peak
IV. According to SWT an excitation corresponding to level $E_3$ between peaks III and IV is expected but not observed.
This problem will be resolved by the DDMRG calculations discussed in the next subsection, which additionally provides
scattering intensities.

Due to the restrictions for the energies mentioned before (cf. Appendix~\ref{sec:AppSW}), a perfect mapping of the
spin-wave spectrum onto the positions of the above three INS peaks is not possible even in the general case of three
exchange parameters. A good approximation to the INS peaks can however be obtained with the two-$J$ model, which is
also compatible with the chemical structure of Fe$_{18}$. A least-squares fit resulted in the optimal parameter
$b_0=0.074$, which yields the energy levels $E_1=3.17$~meV, $E_2=8.67$~meV, $E_3=10.61$~meV, and $E_4=11.84$~meV, which
for $E_1$, $E_2$, and $E_4$ can be compared to the INS energies peak II = 3.0~meV, peak III = 8.5~meV, and peak IV =
12.0~meV. The good agreement suggests that the exchange constants obtained from this analysis should provide excellent
starting values for a more refined analysis.

In fact, from Eq.~(\ref{b0}), for any fixed value of $b_0$ there are two different pairs of exchange constants which
produce identical one-magnon spectra (in the two-$J$ model). The dependence of $b_0$ on the scaling variable
$y=J_2/\sqrt{J_1^2+J_2^2+J_3^2}$ is presented in Fig.~\ref{fig:sp}(b), and the graphical solution of Eq.~(\ref{b0}) for
the optimal parameter $b_0=0.074$ is also shown. The two solutions correspond to $J'/J \approx 3$ and $J'/J \approx
0.3$, which in the following will be referred to as the $J'>J$ and $J'<J$ scenario, respectively. DDMRG considered in
the next subsection allows us to disentangle these two cases.

It is finally mentioned that a good approximation to the experimental spectrum cannot be achieved by a uniform ring or
the one-$J$ model, see also Fig.~\ref{fig:sp}(a). The two-$J$ model is thus both the minimal and appropriate model for
Fe$_{18}$. This is in accordance with similar findings for another, but structurally similar Fe$_{18}$ spin
ring.\cite{Wetal:AngewChemIntEd1997}

\subsection{\label{sec:expana} DDMRG analysis of the experimental data}

In the first part of this section, the magnetic susceptibility and high-energy INS (peaks II-IV) data are analyzed. The
single-ion anisotropy term is neglected and only the Heisenberg Hamiltonian Eq.~(\ref{eq:heisenberg}) is used. In the
second part, using the low-energy INS data (peaks Ia and Ib) the magnitude and influence of the single-ion anisotropy
term Eq.~(\ref{eq:fe18-anis-term}) are investigated.

Using DDMRG\cite{CorrVecRamasesha,CorrVecKuehnerWhite,DDMRG} (see Appendices~\ref{sec:methodsnum} and \ref{sec:appDDMRG}) the INS intensities where first calculated for
the one-$J$ model. The calculated peaks correspond to transitions from the $S = 0$ ground state to excited states with
$S = 1$, and occur approximately at $0.27J$, $2.0J$, $3.6J$, $4.8J$, and $5.5J$. Therefore, as already deduced
from SWT, it is not possible to find a single $J$ for which more than two INS peaks can be reproduced. Furthermore,
simulations of the magnetic susceptibility (see Fig.~\ref{fig:xi}) are in marked discrepancy with experiment. The
one-$J$ model is hence disregarded for interpreting the magnetism in Fe$_{18}$. This model will however be helpful for
the discussion of the general physics in AFM molecular wheels in Sec.~\ref{sec:discussion}.

\begin{figure}
\includegraphics{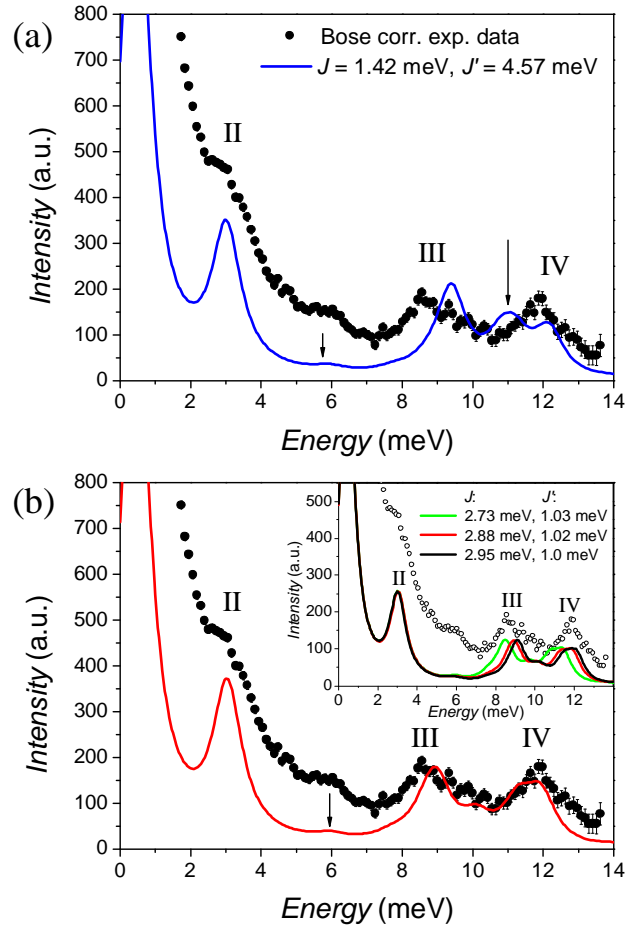}
\caption{\label{fig:fe18-INS-J1J2-Freiburg-Formel}(Color online) INS spectra as obtained by DDMRG calculations (solid
lines) and comparison to the experimental Bose-corrected data (solid and open circles) deduced from the $\lambda$ =
2.26~{\AA}, 1.5~K INS run. (a) Best-fit simulation for the $J'>J$ scenario, with exchange parameters $J=1.42$~meV,
$J'=4.57$~meV. The arrows indicate features in the simulated spectrum discussed in the text. (b) Best-fit simulation
for the $J'<J$ scenario, with exchange parameters $J=2.88$~meV, $J'=1.02$~meV. The inset shows simulated INS spectra
for three slightly different exchange constants used to infer the
errors in the determined exchange constants.}
\end{figure}

Simulations of the magnetic susceptibility for the two-$J$ model (Fig.~\ref{fig:xi}) showed that for both the $J'>J$
and $J'<J$ scenarios the experimental data can be excellently reproduced. The simulations are in fact not very
sensitive to small variations of the exchange constants, as long as their average remains constant.\cite{SSL:PRB01}
Using the information from the SWT calculations and magnetic susceptibility simulations extensive DDMRG calculations
for several $(J,J')$ parameter sets were performed that aimed at fitting the high-energy INS data. The broadening
$\eta$~=~0.5~meV (cf. Appendix~\ref{sec:methodsnum}) was chosen such that in the DDMRG calculation the FWHM of the
Lorentzian peaks approximately corresponds to the experimental resolution (1.1~meV for the $\lambda = 2.26$~\AA~INS
measurements). For small parameter variations (and for not too small a difference between $J$ and $J'$) the following
qualitative results were obtained [cf. Fig.~\ref{fig:fe18-INS-J1J2-Freiburg-Formel}(b)], which for brevity will be
discussed for the $J'<J$ scenario (for the $J'>J$ scenario interchange $J$ and $J'$):

(1) A variation of $J'$ while keeping $J$ constant affects the complete spectrum, but the peak positions depend
approximately linearly on $J'$.

(2) Variation of $J$ with constant $J'$ mainly	shifts the peaks above 6~meV without changing the relative positions of
those peaks, i.e., the "bandwidth" of the high-energy part above 6~meV is not significantly changed. The positions and
heights of the peaks below 6~meV are almost unaffected by variation of $J$.

These findings are in accordance with the SWT results. Using these trends for the $J'>J$ scenario, the best agreement
with experiment was obtained for $J$ = 1.42~meV and $J'$ = 4.57~meV [Fig.~\ref{fig:fe18-INS-J1J2-Freiburg-Formel}(a)].
Five prominent peaks are observed in the DDMRG spectrum at the energies of ca. 0.43, 3.0, 9.4, 10.9, and 12.1~meV
(which can in fact contain some nearly degenerate transitions, which are not resolved due to the finite broadening
$\eta$). The calculated spectrum is compared with the Bose-corrected INS spectrum recorded at a wavelength of
2.26~{\AA} in Fig.~\ref{fig:fe18-INS-J1J2-Freiburg-Formel}(a). The agreement with the experimental peak III could
probably be further optimized by fine-tuning of $J$ and $J'$. However, in the DDMRG spectra an additional peak at about
11~meV is observed, marked by an arrow in Fig.~\ref{fig:fe18-INS-J1J2-Freiburg-Formel}(a), which obviously corresponds
to the spin-wave level $E_3$ discussed in the preceding subsection. It is, however, not seen in the experiment, which
disfavors the $J'>J$ scenario as a model for the magnetism in Fe$_{18}$.

Further efforts therefore concentrated on the $J'<J$ scenario. Here also five peaks are observed in the DDMRG spectra
at ca. 0.43, 3.0, 8.9, 10.1, and 11.6~meV. However, the intensity of the peak at 10.1~meV, which relates to the
spin-wave level $E_3$, is relatively weak. The best agreement of the DDMRG spectra with the experimental high-energy
data was obtained for $J = 2.88$~meV and $J' = 1.02$~meV. The agreement is in fact very good, see
Fig.~\ref{fig:fe18-INS-J1J2-Freiburg-Formel}(b), and in particular considerably better than for the $J'>J$ scenario
[Fig.~\ref{fig:fe18-INS-J1J2-Freiburg-Formel}(a)]. The errors of the determined exchange constants were estimated to
$\sim 0.15$~meV for $J$ and $\sim 0.05$~meV for $J'$. These estimates are based on the variation of the peak positions
for varied parameters [cf. inset to Fig.~\ref{fig:fe18-INS-J1J2-Freiburg-Formel}(b)]. The smaller error for $J'$ stems
from the fact that the experimental position of peak II has been measured more precisely than the positions of peaks
III and IV, and the position of the simulated peak II is mainly determined by this coupling constant. Within these
error bounds, it is possible to match the positions of all experimental high-energy peaks in the DDMRG simulation. The
magnetic susceptibility is also reproduced excellently with these exchange parameters.

The consideration of a non-zero but small temperature in the simulations would not change the positions of the
main peaks, but would change their heights a bit. Additional hot peaks resulting from transitions from excited states
would also appear, but because of the singlet-triplet gap of about 0.3~meV, which should be compared to the temperature
of 1.5~K at which the high-energy INS data were obtained, the ground state population is estimated to be larger than
90~\%. Hence, transitions from the ground state clearly dominate, and comparing the zero-temperature DDMRG and 1.5~K
experimental spectra is justified.

Interestingly, the DDMRG spectra for both scenarios produce a very weak scattering intensity at ca. 6~meV (indicated in
Fig.~\ref{fig:fe18-INS-J1J2-Freiburg-Formel} by arrows), where a weak feature is indeed observed in the experiment. The
analysis of the experimental data in Sec.~\ref{sec:exp} suggested that this intensity is of non-magnetic origin. In
view of the DDMRG results, it could be possible that this feature is, at least in parts, i.e., at low momentum transfer
$Q$, due to magnetic scattering. The present experiments, however, cannot resolve this issue.

\begin{figure}
\includegraphics{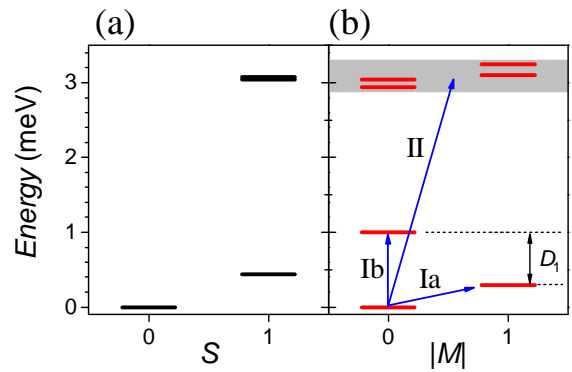}
\caption{
Calculated low-energy spectrum for $J=2.88$~meV, $J'=1.02$~meV as obtained using DMRG with multiple target states.
(a) Energy spectrum for $D=0$ as function of total spin $S$. (b) Energy spectrum for $D=0.03$~meV as function of the
magnitude of the magnetic quantum number $M$, showing the ZFS of the $S=1$ multiplets. For the lowest $S=1$
multiplet the ZFS is determined by $D_1$. The gray box is centered at the center of gravity of the higher-lying
multiplets and has a height corresponding to the experimental resolution (430~$\mu$eV) of the data shown in
Fig.~\ref{fig:INS_3p2A}. The ZFS of the lowest $S=1$ multiplet is $0.7$~meV and much larger than the splitting of the
next two, nearly degenerate multiplets, which is $0.3$~meV, and smaller than the experimental resolution. The observed
cold INS transitions are indicated by arrows.} \label{fig:fe18-2_88-1_02-spectrum-S-D-2}
\end{figure}

In a next step the influence and magnitude of the single-ion anisotropy term, Eq.~(\ref{eq:fe18-anis-term}), are
analyzed. Peaks Ia and Ib, observed at energies of 0.3~meV and 1.0~meV in the $\lambda = 4.2$~{\AA} INS experiment and
in a previous work,\cite{Wal09-fe18} were assigned to the transitions from the $S=0$ ground state to the first excited
$S = 1$ multiplet, which exhibits a Zero-Field Splitting (ZFS) due to the single-ion anisotropy of the Fe$^\text{III}$
ions which can be characterized by an anisotropy constant $D_1$ [see Fig.~\ref{fig:fe18-2_88-1_02-spectrum-S-D-2}(b)].
The relation between $D$ and $D_1$ was estimated in Ref.~\onlinecite{Wal09-fe18} by extrapolating results on smaller
wheels, but the accuracy remained unclear, and therewith the accuracy of the deduced value of $D$ in Fe$_{18}$.
Therefore we used the ALPS DMRG code \cite{ALPS:JMMM07} to calculate the lowest-lying energies for $\widehat{H} = \widehat{H}_H +
\widehat{H}_D$. For $J = 2.88$~meV, $J' = 1.02$~meV, and $D = 0.030$~meV the low-lying energies reproduced excellently
the transition energies observed for peaks Ia and Ib. Based on parameter variations in the calculations and the widths
of the experimentally observed peaks, we estimate the error in $D$ to be $\sim 5$~$\mu$eV.

The comparison of the energy level structure with and without $\widehat{H}_D$ in
Fig.~\ref{fig:fe18-2_88-1_02-spectrum-S-D-2} shows that the anisotropy-induced splitting is largest for the first $S=1$
multiplet but much smaller for the next two (nearly degenerate)
$S=1$ multiplets. This is in agreement with findings on e.g. the
CsFe$_8$ wheel.\cite{Dre10-csfe8ins3} This is a clear 
indication that the effect of the single-ion anisotropy on the higher-lying excitations studied in this work can be
neglected: The position of the center of gravity is nearly unaffected and the splitting is smaller than experimental
resolution. It is mentioned that the center of gravity is not equal to the ``center of INS intensity" since the INS
intensity is generally distributed unequally among the excitations.

To summarize this section, the experimental magnetic susceptibility and INS data for Fe$_{18}$ can very well be
reproduced with the microscopic parameters $J = 2.88(15)$~meV, $J' = 1.02(5)$~meV, and $D = 0.030(5)$~meV [$J =
33(2)$~K, $J' = 11.8(6)$~K, $D = 0.35(6)$~K]. Alternative exchange parameters were extensively searched for but found
to provide inferior agreement with experiment.

\section{\label{sec:discussion} Discussion}

According to the $L \& E$-band picture, the low-temperature excitations in even-membered AFM wheels fall into two
energy regimes with different character. For the wheels Cr$_{8}$ and CsFe$_{8}$ this scenario was confirmed both
experimentally and theoretically in detail,\cite{Wal03-cr8,Dre10-csfe8ins3} but for Fe$_{18}$ the picture has been
incomplete so far. For Fe$_{18}$, the $L$-band excitations at the lowest energies were experimentally investigated in
detail in the previous study Ref.~\onlinecite{Wal09-fe18} by means of low-energy INS (peaks Ia, Ib in
Fig.~\ref{fig:INS_4p2A}) and ultra-low-temperature high-field magnetization and magnetic torque measurements. The
$E$-band or the higher-lying elementary excitations accessible at low temperatures were carefully studied in the
present work. However, different models were used in the interpretation of the low-energy and high-energy data, and
their relation is considered now.

The low-energy experimental data\cite{Wal09-fe18} could be described extremely well by the effective two-sublattice
Hamiltonian
%
\begin{equation} \label{eq:dimermodel}
\widehat{H}_{AB}= j \mathbf{\widehat{S}}_{A}\cdot\mathbf{\widehat{S}}_{B}
+ d \left[(\widehat{S}_{A}^{z})^{2}+(\widehat{S}_{B}^{z})^{2}\right],
\end{equation}
%
where $\mathbf{\widehat{S}}_{A}$ and $\mathbf{\widehat{S}}_{B}$ represent the total spins of length $S_A = S_B = Ns/2$
on each of the two sublattices $A$ and $B$, and with appropriate effective exchange constant $j$ and anisotropy $d$.
This simple two-spin Hamiltonian is suggested by the $L \& E$-band picture, and was established as an effective
low-energy approximation of the uniform ring (one-$J$) model with an additional anisotropy, or $\widehat{H}_{uni} =
\widehat{H}_H + \widehat{H}_D$ with $J_1=J_2=J_3 = J$.\cite{Wal02-wheel-qt} The magnetic parameters are related through
$j = a_1 J$ and $d = b_1 D$, where the coefficients $a_1$ and $b_1$ depend strongly on $N$ and $s$. The underlying
assumption is that the eigenstates of $\widehat{H}_{uni}$ in the low-energy sectors are well approximated by
"quasi-classical" spin states of the form $|S_{A}S_{B}SM\rangle$ (with $S_A = S_B = Ns/2$). Within this space
$\widehat{H}_{uni}$ is equivalent to $\widehat{H}_{AB}$ in first order, yielding $a_1= a_{1}^{AB}$ with $a_{1}^{AB} =
4/N$ and $b_{1}=b_{1}^{AB}$ with $b_{1}^{AB} = (2s-1)/(Ns-1)$. However, quantum corrections modify these parameters
significantly, which can be accounted for by matching the low-energy states to the exact energy spectrum, if the latter
is available, yielding values $a_1^{qm}$ and $b_1^{qm}$.\cite{Wal02-wheel-qt} For $N=18$, $s=5/2$ relevant for
Fe$_{18}$, $a_1^{qm} = 0.2721$ was extracted from QMC calculations,\cite{EnL:PRB06} and $b_1^{qm} \approx 0.07$ was
determined by extrapolating results of rings with lengths of up to $N=10$.\cite{Wal09-fe18,Wal02-wheel-qt} Using these
results $J= 1.64$~meV and $D= 0.026$~meV was inferred for Fe$_{18}$ from the low-energy data in
Ref.~\onlinecite{Wal09-fe18}.

We recalculated the parameters $a_1^{qm}$ and $b_1^{qm}$ as follows: $a_1^{qm}$ was determined from the singlet-triplet
gap of the Heisenberg part of $\widehat{H}_{uni}$ since it can be obtained very accurately by DMRG. For the estimation
of $b_{1}^{qm}$ the spectra of $\widehat{H}_{uni}$ and $\widehat{H}_{AB}$ were compared for different values of $b_{1}$
and $D$ using DMRG and exact diagonalization codes of the ALPS package.\cite{ALPS:JMMM07,Tro:LNCS1999} Our results are
$a_{1}^{qm} = 0.2683(1)$ and $b_{1}^{qm} = 0.063(2)$. Hence, the refined magnetic parameters  $J= 1.64(3)$~meV and $D=
0.029(1)$~meV are obtained, with which $\widehat{H}_{uni}$ describes very well the experimental low-energy excitations
in Fe$_{18}$.

The susceptibility and high-energy INS data, however, cannot be reproduced by a uniform ring model as demonstrated in
Sec.~\ref{sec:expana}. The two-$J$ model was required. From the susceptibility simulations two equally good parameter
sets were found (one with $J'<J$ and one with $J'>J$). Based on susceptibility it was not possible to prefer one set
over the other. Also, spin-wave theory predicted no differences between these two models regarding the excitation
energies. The simulation of the high-energy INS data using DDMRG revealed differences between these two models, and
favored the $J'<J$ model (Fig.~\ref{fig:fe18-INS-J1J2-Freiburg-Formel}).

Also for the three-$J$ model, or $\widehat{H} = \widehat{H}_H + \widehat{H}_D$ with general exchange couplings, the
effective two-sublattice Hamiltonian $\widehat{H}_{AB}$ is obtained in first-order as the effective low-energy
approximation. The modulations in the exchange (and anisotropy) constants along the wheel are effectively averaged out
in the $L$-band states,\cite{Wal01-csfe8torque} suggesting the average exchange constant $J=(J_1+J_2+J_3)/3$. However,
the parameter $a_1$ is strongly modified. Indeed, for the $J'<J$ and $J'>J$ models determined in this work the average
exchange constants are $J = 2.26$~meV and 2.47~meV, while for the uniform ring model $J = 1.64$~meV (all three models
yield identical low-energy spectra, e.g, energies for peaks Ia and Ib). It is interesting to observe that the parameter
$b_1$ is comparatively less affected; for the models here $D= 0.030(5)$~meV is deduced while the uniform ring model
yielded $D = 0.029(1)$~meV, which agree within the errors.

\begin{figure}
\includegraphics{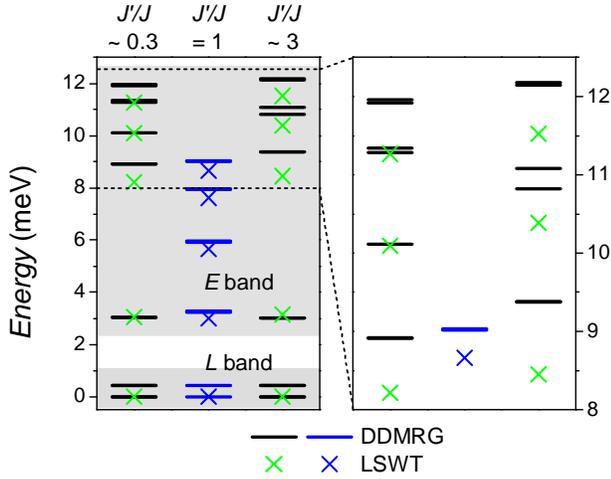}
\caption{\label{fig:collection}
(Color online) Effect of different coupling constants in the Fe$_{18}$ wheel. The left panel shows the ground state and
the excitations in the one-magnon energy sector for the three models $J$ = 2.88~meV, $J'$ = 1.02~meV ($J'/J \approx
0.3$), $J = J'$ = 1.64~meV, and $J$ = 1.42~meV, $J'$ = 4.57~meV ($J'/J' \approx 3$).  The bars represent the energies as
obtained from DDMRG calculations, and the crosses the results of
LSWT. The LSWT excitation energies were calculated using
Eq.~\eqref{ei} (see Appendix~\ref{sec:AppSW}). The right panel
shows a zoom into the 
high-energy region. }
\end{figure}

The energy spectra of the elementary excitations are compared for the $J'<J$, $J'>J$, and uniform ring models in
Fig.~\ref{fig:collection}, where only the Heisenberg parts were considered ($D=0$). Also, the spin-wave energies as
predicted by our SWT (Appendix~\ref{sec:AppSW}) are indicated. Apparently, the lowest-lying triplet, which belongs to the
$L$ band, has identical energy in all three models (as necessitated by experiment). Furthermore, the next-higher lying
triplet state, corresponding to the energy level $E_1$ or peak II, which belongs to the $E$ band, is produced at
nearly identical energies in all models. The main difference is in the further higher-lying states of the $E$ band. In
the one-$J$ model their "bandwidth" of ca. 2.5~meV is obtained approximately correctly, but is predicted at
significantly too low an energy. Hence, the key experimental signature for modulated exchange couplings in Fe$_{18}$ is
the large energy gap between peaks II and III. Notably, for all three models the simple LSWT can reproduce the DDMRG
excitation energies in the $E$ band semi-quantitatively. This strongly suggests that these excitations do indeed
correspond to discrete AFM spin-wave excitations. The lowest-lying singlet-triplet gap or $L$ band is not reproduced by
the LSWT (i.e. is calculated zero energy) by reasons discussed before in Sec.~\ref{sec:swana}.

\begin{figure}
\includegraphics{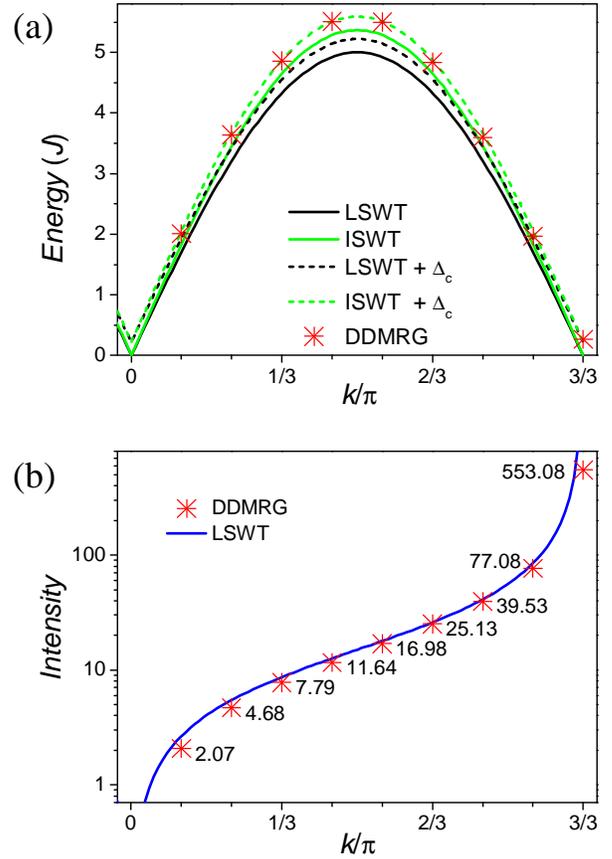}
\caption{\label{fig:disp}
(a) Dispersion of the one-magnon excitations of the uniform Heisenberg ring (one-$J$ model) as obtained by DDMRG
calculations compared with the dispersion relations obtained with LSWT and ISWT (see text for details). The LMSWT and
IMSWT results are not shown as their dispersions are virtually identical to that of LSWT and ISWT, respectively, except
near $k=0,\pm \pi/3$. The $k$ quantum numbers have to be understood relative to the ground state quantum number. (b) Oscillator strength calculated with DDMRG (stars and numbers) compared with the intensity
obtained from LSWT. }
\end{figure}

Finally, the excitations of the uniform Heisenberg ring shall be
considered in detail by comparing the results of (D)DMRG
calculations to those of various SWTs. The discussion parallels
that in Ref.~\onlinecite{Dre10-csfe8ins3} for 
the AFM wheel CsFe$_8$ with $N=8$, $s=5/2$. AFM systems with disordered ground states, such as the AFM wheels,
represent a challenge for any SWT since these start by construction from an ordered ground state. As a result, the
energies of the Goldstone modes are obtained as zero even in finite spin systems, and divergencies appear e.g. in the
magnetization (which actually can be used as an indication of the absence of order in the considered spin
model\cite{And52-swt,IvS:LNP04}). Linear and interacting SWT (LSWT and ISWT) are typical
representatives.\cite{And52-swt,oguchi} These drawbacks can be eliminated by introducing chemical potentials for each
spin center, which result in finite excitation gaps for Goldstone modes (or the singlet-triplet gap) and thereby remove
the divergencies. This branch of SWTs is denoted as modified SWTs,\cite{takahashi} the simplest of which is linear
modified SWT (LMSWT),\cite{sw1,HT:PRB1989,sw3} but also a number of interacting variants exist. Here we use
full-diagonalization interacting modified SWT (IMSWT).\cite{imswt} A conceptually different approach is to introduce
Schwinger bosons and treat the resulting Hamiltonian at the mean-field level (Schwinger-boson mean field
theory),\cite{SBMFT1,SBMFT2} which however yields exactly the same excitation spectrum as LMSWT and is hence not
further considered here. For the $N=8$, $s=5/2$ system it was found that a simple correction of the LSWT spectrum,
called LSWT+$\Delta_c$, gave the best agreement with the exact energies.\cite{Dre10-csfe8ins3} In this approach, the
first-order approximation $\Delta_c = 4J/N$ for the singlet-triplet gap (or $a_1=a_1^{AB}$) is added to the excitation
energies. We will additionally consider ISWT+$\Delta_c$, where the estimated gap is added to the spectrum of ISWT.

\begin{table}
\caption{(D)DMRG and SWT results for the ground-state energy $E_g$, singlet-triplet gap $\Delta$, height of the
dispersion curve, and mean deviation for a $N=18$, $s=5/2$ uniform AFM Heisenberg ring in units of $J$. The numerical
inaccuracy of the DMRG ground state energy is two orders smaller than given.} \label{tab:swt}
\begin{ruledtabular}
		\begin{tabular}{c|cccc}
 & $E_g/J$& $\Delta$/$J$ & height/$J$ & $\chi^2$\\ \hline
(D)DMRG & -129.703  & 0.2683(1) & 5.50(1) & 0\\
LSWT & -128.835 & 0 & 4.924 & 0.0602\\
LMSWT & -129.082 & 0.1228 & 4.926 & 0.0582\\
ISWT & -129.759 & 0 & 5.288 & 0.0246\\
IMSWT & -129.693 & 0.1319 & 5.289 & 0.0246\\
LSWT+$\Delta_C$ & & 0.2222 & 5.146 & 0.0316\\
ISWT+$\Delta_C$ & & 0.2222 & 5.511 & 0.0066
		\end{tabular}
\end{ruledtabular}
\end{table}

For these theories the ground-state and one-magnon energies can be calculated analytically for AFM
wheels.\cite{And52-swt,oguchi,sw1,HT:PRB1989,sw3,imswt,SBMFT1,SBMFT2} The resulting ground-state energies,
singlet-triplet gaps, and maximal energies (= heights of the dispersion curves) are listed in Table~\ref{tab:swt},
which also presents the corresponding values for the one-$J$ model Hamiltonian obtained from (D)DMRG, see appendix
\ref{sec:appDDMRG}. Furthermore the mean deviation of SWT and DDMRG excitation energies or $\chi^2 = \sum_k [ \omega(k)
- \omega_\text{DDMRG}(k)]^2 / ( \frac{1}{2} N J s)$ is given. The excitation energies are also displayed in
Fig.~\ref{fig:disp}(a).

The trends for $N=18$, $s=5/2$ are similar to what has been observed for $N=8$, $s=5/2$:\cite{Dre10-csfe8ins3} For the
ground-state energy IMSWT provides the best approximation to the DMRG result. The singlet-triplet gap cannot be
reproduced by LSWT and ISWT, in contrast to LMSWT and IMSWT. However, their predictions are about a factor two too
small, and in particular significantly poorer than the simple estimate $\Delta_c = 4J/N$. The calculated excitations
for LSWT and LMSWT are nearly identical as seen by similar mean deviations, as well as that for ISWT and IMSWT [LMSWT
and IMSWT dispersion curves were hence not plotted in Fig.~\ref{fig:disp}(a) for clarity]. Thus, as a conclusion,
similar to the situation for $N=8$, $s=5/2$ wheels, IMSWT provides the best results also for $N=18$, $s=5/2$ wheels as
regards the ground-state and $E$-band excitation energies, and the singlet-triplet gap is underestimated by about a
factor of 2. The best description for the excitations is in fact provided by the simple LSWT+$\Delta_c$ and
ISWT+$\Delta_c$ approaches, with a significant advantage of ISWT+$\Delta_c$ (in contrast to the $N=8$ ring, where
LSWT+$\Delta_c$ did better).

Besides the position of the energy levels also the oscillator strengths\cite{Wal02-spindyn} $|\langle
S,M|\widehat{S}^z(k)|S',M'\rangle|^2$ ($\widehat{S}^z(k)=\sum_{j}\exp (-ikj)\widehat{S}_{j}^{z}$) of the transitions from the ground state ($S=0$) to the excited $S'=1$
states were calculated for the uniform $N=18$, $s=5/2$ wheel using DDMRG. The oscillator strength is not identical to
the INS intensity for a specific transition, but both are intimately related.\cite{Wal03-insqdependence} The oscillator
strengths are independent of specific INS-experiment parameters and are thus better suited for general discussions. The
results are summarized in Fig.~\ref{fig:disp}(b). As expected, the oscillator strengths increase with increasing $k$,
becoming maximal at the zone boundary, reminiscent to the behavior of Goldstone modes at the Bragg points $k= \pi$ in
infinite lattices. Interestingly, the oscillator strengths are well described, at least on the logarithmic scale used
in Fig.~\ref{fig:disp}(b), by the predictions of LSWT.\cite{Mue:PRB1982} The LSWT result shows a divergency at the
Bragg point $k= \pi$, which reflects the underlying assumption of a long-range N\'eel-ordered ground state or zero
singlet-triplet gap, as discussed before several times. The good agreement of the DDMRG and LSWT oscillator strength
provides a further strong indication that the elementary excitations in $N=18$, $s=5/2$ AFM wheels are indeed discrete
spin-wave excitations, and for the validity of the $L \& E$-band picture.

\section{\label{sec:conclusions} Conclusions}

In summary, we report a combined experimental and theoretical study of the magnetism in the antiferromagnetic molecular
ferric wheel Fe$_{18}$. It is demonstrated that nowadays advanced experimental and theoretical tools such as inelastic
neutron scattering, (dynamical) density matrix renormalization group techniques, and quantum Monte Carlo together with
established methods such as spin-wave theory permit a detailed characterization of large magnetic molecules with huge
Hilbert space dimensions, such as the molecular ferric wheel Fe$_{18}$. The combined approach allowed us to determine
accurate magnetic parameters as well as to test and rationalize effective models such as the  $L \& E$-band picture,
which are of paramount importance for a deeper understanding of the physics in AFM wheels and quantum-spin clusters in
the mesoscopic regime in general. The concrete result for the Fe$_{18}$ wheel, whose size is situated at the border
between microscopic and macroscopic, is that the higher-lying elementary excitations have the character of discrete
antiferromagnetic spin waves.

\begin{acknowledgments}
The authors thank Eric Jeckelmann and Piet Dargel for discussions about the DDMRG technique. Funding by the Deutsche
Forschungsgemeinschaft (FOR~945, SCHN~615/16-1, WAL~1524/5-1) and the USA National Science Foundation is thankfully
acknowledged. N.B.I. is partially supported by the Bulgarian Science Foundation (Grant No. DO 02-264). This work is
partially based on experiments performed at the Swiss spallation neutron source SINQ, Paul Scherrer Institute,
Villigen, Switzerland.
\end{acknowledgments}

\appendix

\section{\label{sec:methods} Methods}

\subsection{\label{sec:methodsexp} Experimental Methods}

The Fe$_{18}$ material was synthesized according to the procedure described in the literature.\cite{fe18synthese} The
magnetic susceptibility data was recorded on a microcrystalline sample using a SQUID magnetometer (Quantum Design) in
an applied field of 0.5~T. The date were corrected for the contribution of the sample holder. The sample was prepared
by selecting sufficiently many single crystals from the mother liquor, squash them and mount them on the sample holder,
and to cool the sample down as quickly as possible. This procedure typically minimizes contamination by magnetic
impurities. INS spectra were recorded at the time-of-flight disc chopper spectrometer IN5 at the Institute
Laue-Langevin (Grenoble, France) with incoming neutron wavelength $\lambda$ = 4.2~{\AA} and at the direct
time-of-flight spectrometer FOCUS at the Paul Scherrer Institut (Villigen, Switzerland) with incoming neutron
wavelengths $\lambda$ = 3.2 and 2.26~{\AA}. For each INS experiment a fresh non-deuterated powder sample of Fe$_{18}$
was synthesized, and filled in a double-walled hollow aluminum can. The weight of the samples were approximately 0.5
and 1.8~g for the measurements at IN5 and FOCUS, respectively. All data were corrected for detector efficiency with a
measurement of a vanadium standard. At FOCUS also empty can measurements were done and used for empty can corrections.
Experimental resolutions at the elastic line were 165~$\mu$eV in the IN5 experiment, and 430 and 1100~$\mu$eV in the
FOCUS experiments. If not stated otherwise, spectra were summed over all detector banks. Positive energies correspond
to energy loss of the neutron.

\subsection{\label{sec:methodsnum} Numerical Methods}

The INS spectra were calculated as follows. The formula for the differential INS cross section of powder-averaged
isotropic systems\cite{Wal03-insqdependence,Wal05-smix,ts_insformel} reads
\begin{equation} \label{eq:INS-cross-section}
\frac{d^2 \sigma}{d\Omega d\omega}\propto\frac{k'}{k}e^{-2W}F^{2}(Q)S^{zz}(Q,\omega),
\end{equation}
where $\mathbf{Q}=\mathbf{k}-\mathbf{k'}$ is the momentum transfer, $e^{ -2W}$ is the Debye-Waller factor, and $F(Q)$
is the magnetic form factor of Fe$^{III}$ ions. The scattering function $S^{zz}(Q,\omega)$ is defined as
\begin{equation}
S^{zz}(Q,\omega)=\sum_{j,j'}\frac{\sin (QR_{jj'})}{QR_{jj'}}S^{zz}_{jj'}(\omega),
\end{equation}
where $S^{zz}_{jj'}(\omega)$ is at zero temperature given by
\begin{equation} \label{eq:dynam_corr_func}
S^{zz}_{jj'}(\omega)\equiv \sum_{n}\langle 0|\widehat{S}_{j}^{z}|n\rangle\langle
n|\widehat{S}_{j'}^{z}|0\rangle\,\delta (\hbar\omega-E_{n}+E_{0}).
\end{equation}

For the calculation of $S^{zz}_{jj'}(\omega)$ the dynamical density-matrix renormalization group (DDMRG) technique was
used.\cite{CorrVecRamasesha, CorrVecKuehnerWhite,  DDMRG} Details of the calculation are given in
Appendix~\ref{sec:appDDMRG}. Strictly speaking the molecule Fe$_{18}$ is zero-dimensional, but effectively forms a
one-dimensional chain with periodic boundary conditions and is hence suited for DDMRG. However, the calculations are
nevertheless very time-consuming due to the applied periodic boundary conditions. For the calculations on the model
with different coupling constants up to 600 density matrix eigenstates were kept. For the calculations on the uniform
Heisenberg model up to 850 density matrix eigenstates were kept
in order to achieve a smaller broadening $\eta$. The truncated
weight depends on the energies $\hbar\omega$ as well as $\eta$, and ranged between $10^{-7}$ and $10^{-4}$, which is a
very large, but not unusual value for a DDMRG calculation, cf. Ref.~\onlinecite{DDMRGNishimoto}. We checked the results for a different
number of kept density matrix eigenstates and no significant changes of peak positions or heights were observed. For
the direct calculation of the low-lying energy levels the ALPS DMRG code\cite{ALPS:JMMM07} was used, and up to 3000
density matrix eigenstates were kept.

The magnetization of Fe$_{18}$ was evaluated by means of Quantum Monte Carlo (QMC)
calculations\cite{SaK:PRB91,San:PRB99,EnL:PRB06} employing again the ALPS code.\cite{ALPS:JMMM07} Since the underlying
Hilbert space is huge, dimension $(2s+1)^N=101,559,956,668,416$, we used 10,000,000 sweeps for the thermalization and
10,000,000,000 sweeps for the evaluation of the magnetization for every value of the temperature.

All calculations were performed on a BULL supercomputer with 128 cores and 386 GB RAM running ScaleMP vsmp.

\section{Dynamical DMRG \label{sec:appDDMRG}}

The basic steps of the DDMRG technique are briefly discussed.
DDMRG\cite{CorrVecRamasesha,CorrVecKuehnerWhite,DDMRG,DDMRGPeters} is an extension of the standard DMRG
method.\cite{Whi:PRL92,Whi:PRB93,ExS:PRB03,Sch:RMP05} It is a powerful numerical method for the calculation of zero
temperature dynamical correlation functions such as (we set $\hbar=1$ in this section)
\begin{equation}
	G_{A,B}(\omega)=-\frac{1}{\pi}\langle 0|\widehat{A}^{\dagger}\frac{1}{E_{0}+\omega+i\eta-\widehat{H}}\widehat{B}|0\rangle,
\end{equation}
where $|0\rangle$ denotes the ground state with the energy $E_0$. For the comparison to a spectroscopic experimental
method such as inelastic neutron scattering, the important part of this function is the imaginary part (by setting
$\widehat{A}=\widehat{S}_{j}^{z}$ and $\widehat{B}=\widehat{S}_{j'}^{z}$ one obtains the function
$\text{Im}\,G_{A,B}(\omega)=S_{jj'}^{zz}(\omega)$, cf. Sec.~\ref{sec:methodsnum})
\begin{eqnarray}
	\text{Im}\,G_{A,B}(\omega)&=&\frac{1}{\pi}\langle A|\frac{\eta}{(E_{0}+\omega-\widehat{H})^2+\eta^2}|B\rangle\\
	&=&\sum_{n}\langle A|n\rangle\langle n|B\rangle \delta_{\eta}(\omega+E_{0}-E_{n}),
\end{eqnarray}
where $\delta_{\eta}(x)$ is the Lorentzian broadened delta function with $\lim\limits_{\eta \rightarrow
0}\delta_{\eta}(x)=\delta(x)$, and $|A\rangle\equiv \widehat{A}|0\rangle$, $|B\rangle\equiv \widehat{B}|0\rangle$.
$E_n$ denotes the energy eigenvalue belonging to the eigenstate $|n\rangle$. A calculation of the matrix elements
$\langle A|n\rangle$ by directly calculating the excited states $|n\rangle$ using standard DMRG is possible only for
low energies since all energy eigenstates up to the desired state have to be included as target states in forming the
reduced density matrix. Many target states, however, decrease the accuracy of a DMRG calculation.\cite{Whi:PRB93} The
basic idea of the DDMRG method is a reformulation of this equation using the so-called correction vector, which is
defined as\cite{CorrVecKuehnerWhite}
\begin{equation}
	|C(\omega)\rangle = \frac{1}{E_{0}+\omega+i\eta-\widehat{H}}|B\rangle.
\end{equation}
If one splits the correction vector into $|C(\omega)\rangle=|C^{r}(\omega)\rangle+i|C^{i}(\omega)\rangle$ with
\begin{equation} \label{eq:ci-linear-equation}
	|C^{i}(\omega)\rangle = \frac{1}{(E_{0}+\omega-\widehat{H})^2+\eta^2}|B\rangle
\end{equation}
and
\begin{equation}
	|C^{r}(\omega)\rangle=\frac{\widehat{H}-E_{0}-\omega}{\eta}|C^{i}(\omega)\rangle,
\end{equation}
a direct calculation of $\text{Im}\,G_{A,B}(\omega)$ is possible as
\begin{equation}
	\text{Im}\,G_{A,B}(\omega)=\frac{1}{\pi}\langle A|C^{i}(\omega)\rangle.
\end{equation}
$|C^{i}(\omega)\rangle$ is calculated as the solution of a linear equation system within the reduced DMRG basis. The
target states for the reduced density matrix are $|0\rangle$, $|B\rangle$, $|C^{i}(\omega)\rangle$, and
$|C^{r}(\omega)\rangle$. For $A=B$ a different calculation of $\text{Im}\,G_{A,B}(\omega)$ by reformulating
\eqref{eq:ci-linear-equation} as a minimization is also possible\cite{DDMRG} but we found no significant differences
between the two approaches. For solving the linear equation system we use a simple CG algorithm.\cite{Yan:Book2000}
DDMRG is a very time-consuming method because the calculation has to be repeated for every $\omega$. However, since
calculations for different $\omega$ are independent, this method is easy to parallelize.

To gain information about the one-magnon dispersion relation of the uniform chain we have calculated the dynamical
correlation function $S^z(k,\omega)$, which is defined as
\begin{equation}
	S^z(k,\omega)=\sum_{j,j'}e^{i k (j-j')}S_{jj'}^{zz}(\omega)\,,
\end{equation}
($k=2\pi q/N, q=0,1,\dots,17$) and can thus be obtained by simply Fourier transforming the $S_{jj'}^{zz}(\omega)$ data.\cite{DDMRGPeters} A different
way to calculate $S(k,\omega)$ is to set
$\widehat{A}=\widehat{B}=\widehat{S}^{z}(k)\equiv\sum_{j}\exp(-ikj)\widehat{S}_{j}^{z}$
for the calculation of
$\text{Im}\,G_{A,B}(\omega)$.\cite{CorrVecKuehnerWhite}  
However, with this procedure only excitations which contribute
sufficiently to the dynamical correlation function can be detected.
For the transition from the
ground state to the lowest three $S=1$ states it is also possible to directly calculate the transition matrix elements
and the oscillator strengths using standard DMRG (cf. Ref.~\onlinecite{SSBE:PRL2008}). If possible, we have employed and compared
all three approaches to obtain and validate the results shown in Fig.~\ref{fig:disp}. However, the numbers shown in Fig.~\ref{fig:disp} are not numerically exact values. We estimate the relative errors of the oscillator strengths to be smaller than 10 \% in all cases.

\section{Spin-wave theory for the AFM $J_1-J_2-J_3$ Heisenberg Ring \label{sec:AppSW}}

The one-magnon spectrum of the isotropic Heisenberg ring Eq.~(\ref{eq:heisenberg}) or three-$J$ model was considered in
a first-order (next to linear) SWT approximation. Since the unit cell of the model contains three half-integer spins
with $s=5/2$, the Lieb-Mattis theorem implies that the ground state in the case of AFM bonds is a
singlet.\cite{LSM:AP61,LiM:JMP62}

For the SWT calculations it is convenient to introduce spherical coordinates $J_R$, $\theta$, and $\phi$ through the
standard relations:
\begin{eqnarray} \label{spherical}
	\frac{J_1}{J_R}&=& \cos \phi \sin \theta  \equiv x,\nonumber\\
	\frac{J_2}{J_R}&=& \sin \phi \sin \theta \equiv y,\nonumber\\
	\frac{J_3}{J_R}&=& \cos\theta \equiv z,\nonumber\\
	J_R&=& \sqrt{J_1^2+J_2^2+J_3^2},
\end{eqnarray}
where $0\leq \phi \leq \frac{\pi}{2}$ and $0\leq \theta \leq \pi/2$ for AFM exchange constants. The radial coordinate
$J_R$ appears as an overall factor in the Hamiltonian, Eq.~(\ref{eq:heisenberg}), and sets the energy scale. The spin
operators were parameterized as follows:
\begin{eqnarray}
\label{pq}
\widehat{S}_{l,\alpha}^z&=&\cos \Phi \left[
s+\frac{1}{2}-\frac{1}{2}(\widehat{p}_{l,\alpha}^2+\widehat{q}_{l,\alpha}^2)
\right]-\sin \Phi\,\sqrt{S}\,\widehat{q}_{l,\alpha},
\nonumber\\
\widehat{S}_{l,\alpha}^x&=&\sin \Phi \left[
s+\frac{1}{2}-\frac{1}{2}(\widehat{p}_{l,\alpha}^2+\widehat{q}_{l,\alpha}^2)\right]
+\cos \Phi\,\sqrt{S}\,\widehat{q}_{l,\alpha},
\nonumber\\
\widehat{S}_{l,\alpha}^y&=&\sqrt{s}\,\widehat{p}_{l,\alpha},
\end{eqnarray}
where $\alpha = 1, 2, 3$ and $l = 1, 2, \dots, L$. The angle $\Phi$ alternatively takes the two values $0$ and $\pi$ on
the lattice sites along the ring. The coordinate $\widehat{q}_{l,\alpha}$ and momentum $\widehat{p}_{l,\alpha}$ satisfy
the usual commutation relations $[\widehat{q}_{l,\alpha},\widehat{p}_{l',\beta}]= i \delta_{l,l'}
\delta_{\alpha,\beta}$.

In terms of the Fourier transforms
\begin{equation}
(\widehat{q}_{k,\alpha},\widehat{p}_{k,\alpha})=\frac{1}{\sqrt{L}}\sum_{l=1}^{L}
(\widehat{q}_{l,\alpha},\widehat{p}_{l,\alpha})\exp{(-ikl)}
\end{equation}
the linear spin-wave theory (LSWT) Hamiltonian reads
\begin{equation}
\widehat{H}_{LSWT}=E_g^{'}+\frac{J_R s}{2}\sum_k\left(
\mathbf{\widehat{p}}_{-k}\cdot\mathbf{M}_{k}\cdot\mathbf{\widehat{p}}_{k}+
\mathbf{\widehat{q}}_{-k}\cdot\mathbf{N}_{k}\cdot\mathbf{\widehat{q}}_{k}\right),
\end{equation}
where $E_g^{'}=-J_R\left(x+y+z\right) s\left( s+1\right)L$,
$\mathbf{\widehat{q}}_k=(\widehat{q}_{k,1},\widehat{q}_{k,2},\widehat{q}_{k,3})$ and
$\mathbf{\widehat{p}}_k=(\widehat{p}_{k,1},\widehat{p}_{k,2},\widehat{p}_{k,3})$. The Hermitian matrices $\mathbf{M}_k$
and $\mathbf{N}_k$ read
\begin{equation}
M_k^{\alpha\beta}=
\left(
\begin{array}{ccc}
x+z & x & z\exp (-3ik) \\
x & x+y & y \\
z\exp (3ik) & y & y+z
\end{array} \right),
\end{equation}
\begin{equation}
N_k^{\alpha\beta}= \left(
\begin{array}{ccc}
x+z & -x & - z\exp (-3ik) \\
-x & x+y & -y \\
-z\exp (3ik) & -y & y+z
\end{array} \right).
\end{equation}
The Hamilton equations of motion for the vectors $\mathbf{\widehat{q}}_k$ and $\mathbf{\widehat{p}}_k$ take the form
\begin{eqnarray}
\frac{d\mathbf{\widehat{q}}_k}{dt} &=& J_R s\mathbf{M}_k\mathbf{\widehat{p}}_k,
\\
\frac{d\mathbf{\widehat{p}}_k}{dt} &=& -J_R s\mathbf{N}_k\mathbf{\widehat{q}}_k,
\end{eqnarray}
which yield three spin-wave branches with dispersions $\omega_{\alpha}(k)$ defined by the secular cubic equation
\begin{equation}
\det
\left(\mathbf{M}_k\cdot\mathbf{N}_k - \frac{\omega_{\alpha}(k)^2}{ J_R^2 s^2}\mathbf{I}\right)=0,
\end{equation}
where $\mathbf{I}$ is the unit ($3\times 3$) matrix. In diagonal form the spin-wave Hamiltonian $\widehat{H}_{LSWT}$
then becomes
\begin{equation}
\label{swh}
\widehat{H}_{LSWT}=E_g+\sum_{\alpha=1}^3\sum_k \omega_{\alpha}(k)\widehat{n}_{k,\alpha},
\end{equation}
where $\widehat{n}_{k,\alpha}$ is the occupation number operator of the $(k,\alpha)$ state and $E_g$ is  the
ground-state energy:
\begin{equation}
\label{swe0}
E_g=- \frac{J_1+J_2+J_3}{3} N s\left(s+1\right) + \frac{1}{2}\sum_{\alpha=1}^3\sum_k \omega_{\alpha}(k).
\end{equation}

It is useful to present the cubic equation for $\omega_{\alpha}(k)$ in the following dimensionless form
\begin{equation}
\label{y}
\xi^3+a_2\xi^2+a_1\xi+a_0=0,
\end{equation}
with
\begin{eqnarray}
\label{aa}
 a_0&=&-\frac{4x^2y^2z^2}{(xy+yz+zx)^3}\sin^2
(3k),\nonumber\\
a_1&=&1,\nonumber\\
a_2&=&-2.
\end{eqnarray}
The spin-wave energies are related to the three roots $\xi_{k,\alpha}$ of Eq.~(\ref{y}) by the expressions
\begin{equation}
\label{omega}
\omega_{\alpha}(k)=
\sqrt{J_1J_2+J_2J_3+J_3J_1} \; s \left(1+\frac{R}{s}\right) \sqrt{\xi_{k,\alpha}}.
\end{equation}
Notice that in this expression the first-order correction to the one-magnon energies was introduced through Oguchi's
renormalization factor $R$.\cite{oguchi} For the uniform ring or one-$J$ model holds  $R=1/2-1/\pi\approx 0.1817$. In
the general case, $R$ is expected to be a smooth regular function of the parameters $\theta$ and $\phi$. In principle,
one can calculate this function in the next-order SWT. However, the correction is small (few $\%$)and is hence not
employed here. For the three-$J$ model of Fe$_{18}$, the spin-wave excitations are distributed in five energy levels
defined as follows:
\begin{eqnarray}
\label{ei}
L_0&=&\omega_{1}(0) = \omega_{1}(\pi/3),\nonumber\\
E_1&=&\omega_{1}(\pm \pi/9) = \omega_{1}(\pm 2\pi/9),\nonumber\\
E_2&=&\omega_{2}(\pm \pi/9) = \omega_{2}(\pm 2\pi/9),\nonumber\\
E_3&=&\omega_{2}(0) = \omega_{3}(0) = \omega_{2}(\pi/3) = \omega_{3}(\pi/3),\nonumber\\
E_4&=&\omega_{3}(\pm \pi/9) = \omega_{3}(\pm 2\pi/9).
\end{eqnarray}
The degeneracies are dictated by the spatial $C_6$ symmetry plus the underlying bipartite sublattice structure.

Since the coefficient $a_0$ in Eq.~(\ref{y}) vanishes for $k=0$ and $\pm \pi/3$, the roots at these special points of
the Brillouin zone can easily be determined to $\xi_{1}=0$, $\xi_{2,3}=1$. In terms of the energy levels one hence
finds $L_0=0$ and
\begin{equation}
\label{e3}
E_3 = s\left(1+\frac{R}{s}\right)\sqrt{J_1J_2+J_2J_3+J_3J_1},
\end{equation}
and the energy $E_3$ plays the role of an overall prefactor in the dispersion of the magnon excitations since
\begin{equation}\label{swe}
\omega_{\alpha}(k) = E_3 \sqrt{\xi_{k,\alpha}}.
\end{equation}
Actually, $E_3$ absorbs the explicit dependence on the parameters $J_1$, $J_2$ and $J_3$ apart from the dependence of
$\xi_{k,\alpha}$ through $a_0$.

Useful relations between the energy levels $E_1$, $E_2$, $E_3$, and $E_4$ follow from Vi\`{e}te's formulas at
$k=\pi/9$: $\xi_{k,1}+\xi_{k,2}+\xi_{k,3}=-a_2=2$, $\xi_{k,1}\xi_{k,2}+\xi_{k,2}\xi_{k,3}+\xi_{k,3}\xi_{k,1}=a_1=1$,
and $\xi_{k,1}\xi_{k,2}\xi_{k,3}=b_0$, where $b_0\equiv-a_0(\pi/9)$. With Eqs.~(\ref{ei}) one obtains
\begin{eqnarray}
\label{viete}
2 E_3^2&=&E_1^2 + E_2^2 + E_4^2,\nonumber\\
E_3^4 &=& E_1^2 E_2^2 +  E_2^2 E_4^2 + E_4^2 E_1^2,\nonumber\\
b_0 E_3^6 &=& E_1^2 E_2^2 E_4^2.
\end{eqnarray}
Equations~(\ref{e3}) and (\ref{viete}) are very handy to investigate the trends in the one-magnon spectrum with varying
exchange constants $J_1$, $J_2$ and $J_3$, and led to a number of general conclusions which are useful for analyzing
the INS peaks, see Sec.~\ref{sec:swana}.

\bibliography{fe18-ref}

\begin{thebibliography}{66}
\expandafter\ifx\csname natexlab\endcsname\relax\def\natexlab#1{#1}\fi
\expandafter\ifx\csname bibnamefont\endcsname\relax
  \def\bibnamefont#1{#1}\fi
\expandafter\ifx\csname bibfnamefont\endcsname\relax
  \def\bibfnamefont#1{#1}\fi
\expandafter\ifx\csname citenamefont\endcsname\relax
  \def\citenamefont#1{#1}\fi
\expandafter\ifx\csname url\endcsname\relax
  \def\url#1{\texttt{#1}}\fi
\expandafter\ifx\csname urlprefix\endcsname\relax\def\urlprefix{URL }\fi
\providecommand{\bibinfo}[2]{#2}
\providecommand{\eprint}[2][]{\url{#2}}

\bibitem[{\citenamefont{B{\"a}rwinkel et~al.}(2000)\citenamefont{B{\"a}rwinkel,
  Schmidt, and Schnack}}]{BSS:JMMM00:B}
\bibinfo{author}{\bibfnamefont{K.}~\bibnamefont{B{\"a}rwinkel}},
  \bibinfo{author}{\bibfnamefont{H.-J.} \bibnamefont{Schmidt}},
  \bibnamefont{and} \bibinfo{author}{\bibfnamefont{J.}~\bibnamefont{Schnack}},
  \bibinfo{journal}{J. Magn. Magn. Mater.} \textbf{\bibinfo{volume}{220}},
  \bibinfo{pages}{227} (\bibinfo{year}{2000}).

\bibitem[{\citenamefont{Taft et~al.}(1994)\citenamefont{Taft, Delfs,
  Papaefthymiou, Foner, Gatteschi, and Lippard}}]{mag_af_ring}
\bibinfo{author}{\bibfnamefont{K.~L.} \bibnamefont{Taft}},
  \bibinfo{author}{\bibfnamefont{C.~D.} \bibnamefont{Delfs}},
  \bibinfo{author}{\bibfnamefont{G.~C.} \bibnamefont{Papaefthymiou}},
  \bibinfo{author}{\bibfnamefont{S.}~\bibnamefont{Foner}},
  \bibinfo{author}{\bibfnamefont{D.}~\bibnamefont{Gatteschi}},
  \bibnamefont{and} \bibinfo{author}{\bibfnamefont{S.~J.}
  \bibnamefont{Lippard}}, \bibinfo{journal}{J.~Am. Chem. Soc.}
  \textbf{\bibinfo{volume}{116}}, \bibinfo{pages}{823} (\bibinfo{year}{1994}).

\bibitem[{\citenamefont{Saalfrank et~al.}(1997)\citenamefont{Saalfrank, Bernt,
  Uller, and Hampel}}]{SBUH:AngewChemIntEd1997}
\bibinfo{author}{\bibfnamefont{R.~W.} \bibnamefont{Saalfrank}},
  \bibinfo{author}{\bibfnamefont{I.}~\bibnamefont{Bernt}},
  \bibinfo{author}{\bibfnamefont{E.}~\bibnamefont{Uller}}, \bibnamefont{and}
  \bibinfo{author}{\bibfnamefont{F.}~\bibnamefont{Hampel}},
  \bibinfo{journal}{Angew. Chem. Int. Ed.} \textbf{\bibinfo{volume}{36}},
  \bibinfo{pages}{2482} (\bibinfo{year}{1997}).

\bibitem[{\citenamefont{Watton et~al.}(1997)\citenamefont{Watton, Fuhrmann,
  Pence, Lippard, Caneschi, Cornia, and Abbati}}]{Wetal:AngewChemIntEd1997}
\bibinfo{author}{\bibfnamefont{S.~P.} \bibnamefont{Watton}},
  \bibinfo{author}{\bibfnamefont{P.}~\bibnamefont{Fuhrmann}},
  \bibinfo{author}{\bibfnamefont{L.~E.} \bibnamefont{Pence}},
  \bibinfo{author}{\bibfnamefont{S.~J.} \bibnamefont{Lippard}},
  \bibinfo{author}{\bibfnamefont{A.}~\bibnamefont{Caneschi}},
  \bibinfo{author}{\bibfnamefont{A.}~\bibnamefont{Cornia}}, \bibnamefont{and}
  \bibinfo{author}{\bibfnamefont{G.~L.} \bibnamefont{Abbati}},
  \bibinfo{journal}{Angew. Chem. Int. Ed.} \textbf{\bibinfo{volume}{36}},
  \bibinfo{pages}{2774} (\bibinfo{year}{1997}).

\bibitem[{\citenamefont{Chiolero and Loss}(1998)}]{Chi98}
\bibinfo{author}{\bibfnamefont{A.}~\bibnamefont{Chiolero}} \bibnamefont{and}
  \bibinfo{author}{\bibfnamefont{D.}~\bibnamefont{Loss}},
  \bibinfo{journal}{Phys. Rev. Lett.} \textbf{\bibinfo{volume}{80}},
  \bibinfo{pages}{169} (\bibinfo{year}{1998}).

\bibitem[{\citenamefont{Abbati et~al.}(2000)\citenamefont{Abbati, Caneschi,
  Cornia, Fabretti, and Gatteschi}}]{ACCFG:InorgChimActa2000}
\bibinfo{author}{\bibfnamefont{G.}~\bibnamefont{Abbati}},
  \bibinfo{author}{\bibfnamefont{A.}~\bibnamefont{Caneschi}},
  \bibinfo{author}{\bibfnamefont{A.}~\bibnamefont{Cornia}},
  \bibinfo{author}{\bibfnamefont{A.}~\bibnamefont{Fabretti}}, \bibnamefont{and}
  \bibinfo{author}{\bibfnamefont{D.}~\bibnamefont{Gatteschi}},
  \bibinfo{journal}{Inorganica Chimica Acta} \textbf{\bibinfo{volume}{297}},
  \bibinfo{pages}{291 } (\bibinfo{year}{2000}).

\bibitem[{\citenamefont{van Slageren et~al.}(2002)\citenamefont{van Slageren,
  Sessoli, Gatteschi, Smith, Helliwell, Winpenny, Cornia, Barra, Jansen,
  Rentschler et~al.}}]{vSetal:ChemEurJ2002}
\bibinfo{author}{\bibfnamefont{J.}~\bibnamefont{van Slageren}},
  \bibinfo{author}{\bibfnamefont{R.}~\bibnamefont{Sessoli}},
  \bibinfo{author}{\bibfnamefont{D.}~\bibnamefont{Gatteschi}},
  \bibinfo{author}{\bibfnamefont{A.~A.} \bibnamefont{Smith}},
  \bibinfo{author}{\bibfnamefont{M.}~\bibnamefont{Helliwell}},
  \bibinfo{author}{\bibfnamefont{R.~E.~P.} \bibnamefont{Winpenny}},
  \bibinfo{author}{\bibfnamefont{A.}~\bibnamefont{Cornia}},
  \bibinfo{author}{\bibfnamefont{A.-L.} \bibnamefont{Barra}},
  \bibinfo{author}{\bibfnamefont{A.~G.~M.} \bibnamefont{Jansen}},
  \bibinfo{author}{\bibfnamefont{E.}~\bibnamefont{Rentschler}},
  \bibnamefont{et~al.}, \bibinfo{journal}{Chem. Eur. J.}
  \textbf{\bibinfo{volume}{8}}, \bibinfo{pages}{277} (\bibinfo{year}{2002}).

\bibitem[{\citenamefont{B\"arwinkel et~al.}(2003)\citenamefont{B\"arwinkel,
  Hage, Schmidt, and Schnack}}]{BHS:PRB03}
\bibinfo{author}{\bibfnamefont{K.}~\bibnamefont{B\"arwinkel}},
  \bibinfo{author}{\bibfnamefont{P.}~\bibnamefont{Hage}},
  \bibinfo{author}{\bibfnamefont{H.-J.} \bibnamefont{Schmidt}},
  \bibnamefont{and} \bibinfo{author}{\bibfnamefont{J.}~\bibnamefont{Schnack}},
  \bibinfo{journal}{Phys. Rev. B} \textbf{\bibinfo{volume}{68}},
  \bibinfo{pages}{054422} (\bibinfo{year}{2003}).

\bibitem[{\citenamefont{Waldmann}(2005)}]{Wal05-gridreview}
\bibinfo{author}{\bibfnamefont{O.}~\bibnamefont{Waldmann}},
  \bibinfo{journal}{Coordin. Chem. Rev.} \textbf{\bibinfo{volume}{249}},
  \bibinfo{pages}{2550} (\bibinfo{year}{2005}).

\bibitem[{\citenamefont{Waldmann et~al.}(2003)\citenamefont{Waldmann, Guidi,
  Carretta, Mondelli, and Dearden}}]{Wal03-cr8}
\bibinfo{author}{\bibfnamefont{O.}~\bibnamefont{Waldmann}},
  \bibinfo{author}{\bibfnamefont{T.}~\bibnamefont{Guidi}},
  \bibinfo{author}{\bibfnamefont{S.}~\bibnamefont{Carretta}},
  \bibinfo{author}{\bibfnamefont{C.}~\bibnamefont{Mondelli}}, \bibnamefont{and}
  \bibinfo{author}{\bibfnamefont{A.~L.} \bibnamefont{Dearden}},
  \bibinfo{journal}{Phys. Rev. Lett.} \textbf{\bibinfo{volume}{91}},
  \bibinfo{pages}{237202} (\bibinfo{year}{2003}).

\bibitem[{\citenamefont{Micotti et~al.}(2006)\citenamefont{Micotti, Furukawa,
  Kumagai, Carretta, Lascialfari, Borsa, Timco, and Winpenny}}]{MFK:PRL06}
\bibinfo{author}{\bibfnamefont{E.}~\bibnamefont{Micotti}},
  \bibinfo{author}{\bibfnamefont{Y.}~\bibnamefont{Furukawa}},
  \bibinfo{author}{\bibfnamefont{K.}~\bibnamefont{Kumagai}},
  \bibinfo{author}{\bibfnamefont{S.}~\bibnamefont{Carretta}},
  \bibinfo{author}{\bibfnamefont{A.}~\bibnamefont{Lascialfari}},
  \bibinfo{author}{\bibfnamefont{F.}~\bibnamefont{Borsa}},
  \bibinfo{author}{\bibfnamefont{G.~A.} \bibnamefont{Timco}}, \bibnamefont{and}
  \bibinfo{author}{\bibfnamefont{R.~E.~P.} \bibnamefont{Winpenny}},
  \bibinfo{journal}{Phys. Rev. Lett.} \textbf{\bibinfo{volume}{97}},
  \bibinfo{pages}{267204} (\bibinfo{year}{2006}).

\bibitem[{\citenamefont{Hoshino et~al.}(2009)\citenamefont{Hoshino, Nakano,
  Nojiri, Wernsdorfer, and Oshio}}]{HNN:JACS09}
\bibinfo{author}{\bibfnamefont{N.}~\bibnamefont{Hoshino}},
  \bibinfo{author}{\bibfnamefont{M.}~\bibnamefont{Nakano}},
  \bibinfo{author}{\bibfnamefont{H.}~\bibnamefont{Nojiri}},
  \bibinfo{author}{\bibfnamefont{W.}~\bibnamefont{Wernsdorfer}},
  \bibnamefont{and} \bibinfo{author}{\bibfnamefont{H.}~\bibnamefont{Oshio}},
  \bibinfo{journal}{J. Am. Chem. Soc.} \textbf{\bibinfo{volume}{131}},
  \bibinfo{pages}{15100} (\bibinfo{year}{2009}).

\bibitem[{\citenamefont{King et~al.}(2006)\citenamefont{King, Stamatatos,
  Abboud, and Christou}}]{fe18synthese}
\bibinfo{author}{\bibfnamefont{P.}~\bibnamefont{King}},
  \bibinfo{author}{\bibfnamefont{T.~C.} \bibnamefont{Stamatatos}},
  \bibinfo{author}{\bibfnamefont{K.~A.} \bibnamefont{Abboud}},
  \bibnamefont{and} \bibinfo{author}{\bibfnamefont{G.}~\bibnamefont{Christou}},
  \bibinfo{journal}{Angew. Chem. Int. Ed.} \textbf{\bibinfo{volume}{45}},
  \bibinfo{pages}{7379} (\bibinfo{year}{2006}).

\bibitem[{\citenamefont{Waldmann et~al.}(2009)\citenamefont{Waldmann,
  Stamatatos, Christou, G\"udel, Sheikin, and Mutka}}]{Wal09-fe18}
\bibinfo{author}{\bibfnamefont{O.}~\bibnamefont{Waldmann}},
  \bibinfo{author}{\bibfnamefont{T.~C.} \bibnamefont{Stamatatos}},
  \bibinfo{author}{\bibfnamefont{G.}~\bibnamefont{Christou}},
  \bibinfo{author}{\bibfnamefont{H.~U.} \bibnamefont{G\"udel}},
  \bibinfo{author}{\bibfnamefont{I.}~\bibnamefont{Sheikin}}, \bibnamefont{and}
  \bibinfo{author}{\bibfnamefont{H.}~\bibnamefont{Mutka}},
  \bibinfo{journal}{Phys. Rev. Lett.} \textbf{\bibinfo{volume}{102}},
  \bibinfo{pages}{157202} (\bibinfo{year}{2009}).

\bibitem[{\citenamefont{Schnack and Luban}(2000)}]{ScL:PRB00}
\bibinfo{author}{\bibfnamefont{J.}~\bibnamefont{Schnack}} \bibnamefont{and}
  \bibinfo{author}{\bibfnamefont{M.}~\bibnamefont{Luban}},
  \bibinfo{journal}{Phys. Rev. B} \textbf{\bibinfo{volume}{63}},
  \bibinfo{pages}{014418} (\bibinfo{year}{2000}).

\bibitem[{\citenamefont{Waldmann}(2002{\natexlab{a}})}]{Wal02-spindyn}
\bibinfo{author}{\bibfnamefont{O.}~\bibnamefont{Waldmann}},
  \bibinfo{journal}{Phys. Rev. B} \textbf{\bibinfo{volume}{65}},
  \bibinfo{pages}{024424} (\bibinfo{year}{2002}{\natexlab{a}}).

\bibitem[{\citenamefont{Schnack et~al.}(2001)\citenamefont{Schnack, Luban, and
  Modler}}]{SLM:EPL01}
\bibinfo{author}{\bibfnamefont{J.}~\bibnamefont{Schnack}},
  \bibinfo{author}{\bibfnamefont{M.}~\bibnamefont{Luban}}, \bibnamefont{and}
  \bibinfo{author}{\bibfnamefont{R.}~\bibnamefont{Modler}},
  \bibinfo{journal}{Europhys. Lett.} \textbf{\bibinfo{volume}{56}},
  \bibinfo{pages}{863} (\bibinfo{year}{2001}).

\bibitem[{\citenamefont{Dreiser et~al.}(2010)\citenamefont{Dreiser, Waldmann,
  Dobe, Carver, Ochsenbein, Sieber, G\"udel, van Duijn, Taylor, and
  Podlesnyak}}]{Dre10-csfe8ins3}
\bibinfo{author}{\bibfnamefont{J.}~\bibnamefont{Dreiser}},
  \bibinfo{author}{\bibfnamefont{O.}~\bibnamefont{Waldmann}},
  \bibinfo{author}{\bibfnamefont{C.}~\bibnamefont{Dobe}},
  \bibinfo{author}{\bibfnamefont{G.}~\bibnamefont{Carver}},
  \bibinfo{author}{\bibfnamefont{S.~T.} \bibnamefont{Ochsenbein}},
  \bibinfo{author}{\bibfnamefont{A.}~\bibnamefont{Sieber}},
  \bibinfo{author}{\bibfnamefont{H.~U.} \bibnamefont{G\"udel}},
  \bibinfo{author}{\bibfnamefont{J.}~\bibnamefont{van Duijn}},
  \bibinfo{author}{\bibfnamefont{J.}~\bibnamefont{Taylor}}, \bibnamefont{and}
  \bibinfo{author}{\bibfnamefont{A.}~\bibnamefont{Podlesnyak}},
  \bibinfo{journal}{Phys. Rev. B} \textbf{\bibinfo{volume}{81}},
  \bibinfo{pages}{024408} (\bibinfo{year}{2010}).

\bibitem[{\citenamefont{Waldmann}(2007)}]{Wal07-swtfe30}
\bibinfo{author}{\bibfnamefont{O.}~\bibnamefont{Waldmann}},
  \bibinfo{journal}{Phys. Rev. B} \textbf{\bibinfo{volume}{75}},
  \bibinfo{pages}{012415} (\bibinfo{year}{2007}).

\bibitem[{\citenamefont{Schnalle and Schnack}(2010)}]{ScS:IRPC10}
\bibinfo{author}{\bibfnamefont{R.}~\bibnamefont{Schnalle}} \bibnamefont{and}
  \bibinfo{author}{\bibfnamefont{J.}~\bibnamefont{Schnack}},
  \bibinfo{journal}{Int. Rev. Phys. Chem.} \textbf{\bibinfo{volume}{29}},
  \bibinfo{pages}{403} (\bibinfo{year}{2010}).

\bibitem[{\citenamefont{Delfs et~al.}(1993)\citenamefont{Delfs, Gatteschi,
  Pardi, Sessoli, Wieghardt, and Hanke}}]{DGP:IC93}
\bibinfo{author}{\bibfnamefont{C.}~\bibnamefont{Delfs}},
  \bibinfo{author}{\bibfnamefont{D.}~\bibnamefont{Gatteschi}},
  \bibinfo{author}{\bibfnamefont{L.}~\bibnamefont{Pardi}},
  \bibinfo{author}{\bibfnamefont{R.}~\bibnamefont{Sessoli}},
  \bibinfo{author}{\bibfnamefont{K.}~\bibnamefont{Wieghardt}},
  \bibnamefont{and} \bibinfo{author}{\bibfnamefont{D.}~\bibnamefont{Hanke}},
  \bibinfo{journal}{Inorg. Chem.} \textbf{\bibinfo{volume}{32}},
  \bibinfo{pages}{3099} (\bibinfo{year}{1993}).

\bibitem[{\citenamefont{Waldmann}(2000)}]{Wal00-sym}
\bibinfo{author}{\bibfnamefont{O.}~\bibnamefont{Waldmann}},
  \bibinfo{journal}{Phys. Rev. B} \textbf{\bibinfo{volume}{61}},
  \bibinfo{pages}{6138} (\bibinfo{year}{2000}).

\bibitem[{\citenamefont{Pilawa et~al.}(2005)\citenamefont{Pilawa, Boffinger,
  Keilhauer, Leppin, Odenwald, Wendl, Berthier, and Horvatic}}]{Pil05}
\bibinfo{author}{\bibfnamefont{B.}~\bibnamefont{Pilawa}},
  \bibinfo{author}{\bibfnamefont{R.}~\bibnamefont{Boffinger}},
  \bibinfo{author}{\bibfnamefont{I.}~\bibnamefont{Keilhauer}},
  \bibinfo{author}{\bibfnamefont{R.}~\bibnamefont{Leppin}},
  \bibinfo{author}{\bibfnamefont{I.}~\bibnamefont{Odenwald}},
  \bibinfo{author}{\bibfnamefont{W.}~\bibnamefont{Wendl}},
  \bibinfo{author}{\bibfnamefont{C.}~\bibnamefont{Berthier}}, \bibnamefont{and}
  \bibinfo{author}{\bibfnamefont{M.}~\bibnamefont{Horvatic}},
  \bibinfo{journal}{Phys. Rev. B} \textbf{\bibinfo{volume}{71}},
  \bibinfo{pages}{184419} (\bibinfo{year}{2005}).

\bibitem[{\citenamefont{Sandvik and Kurkij\"arvi}(1991)}]{SaK:PRB91}
\bibinfo{author}{\bibfnamefont{A.~W.} \bibnamefont{Sandvik}} \bibnamefont{and}
  \bibinfo{author}{\bibfnamefont{J.}~\bibnamefont{Kurkij\"arvi}},
  \bibinfo{journal}{Phys. Rev. B} \textbf{\bibinfo{volume}{43}},
  \bibinfo{pages}{5950} (\bibinfo{year}{1991}).

\bibitem[{\citenamefont{Sandvik}(1999)}]{San:PRB99}
\bibinfo{author}{\bibfnamefont{A.~W.} \bibnamefont{Sandvik}},
  \bibinfo{journal}{Phys. Rev. B} \textbf{\bibinfo{volume}{59}},
  \bibinfo{pages}{R14157} (\bibinfo{year}{1999}).

\bibitem[{\citenamefont{Albuquerque et~al.}(2007)\citenamefont{Albuquerque,
  Alet, Corboz, Dayal, Feiguin, Fuchs, Gamper, Gull, G{\"u}rtler, Honecker
  et~al.}}]{ALPS:JMMM07}
\bibinfo{author}{\bibfnamefont{A.}~\bibnamefont{Albuquerque}},
  \bibinfo{author}{\bibfnamefont{F.}~\bibnamefont{Alet}},
  \bibinfo{author}{\bibfnamefont{P.}~\bibnamefont{Corboz}},
  \bibinfo{author}{\bibfnamefont{P.}~\bibnamefont{Dayal}},
  \bibinfo{author}{\bibfnamefont{A.}~\bibnamefont{Feiguin}},
  \bibinfo{author}{\bibfnamefont{S.}~\bibnamefont{Fuchs}},
  \bibinfo{author}{\bibfnamefont{L.}~\bibnamefont{Gamper}},
  \bibinfo{author}{\bibfnamefont{E.}~\bibnamefont{Gull}},
  \bibinfo{author}{\bibfnamefont{S.}~\bibnamefont{G{\"u}rtler}},
  \bibinfo{author}{\bibfnamefont{A.}~\bibnamefont{Honecker}},
  \bibnamefont{et~al.} (\bibinfo{collaboration}{ALPS collaboration}),
  \bibinfo{journal}{J. Magn. Magn. Mater.} \textbf{\bibinfo{volume}{310}},
  \bibinfo{pages}{1187 } (\bibinfo{year}{2007}), \bibinfo{note}{see also
  \url{http://alps.comp-phys.org}}.

\bibitem[{\citenamefont{White}(1993)}]{Whi:PRB93}
\bibinfo{author}{\bibfnamefont{S.~R.} \bibnamefont{White}},
  \bibinfo{journal}{Phys. Rev. B} \textbf{\bibinfo{volume}{48}},
  \bibinfo{pages}{10345} (\bibinfo{year}{1993}).

\bibitem[{\citenamefont{Schollw{\"o}ck}(2005)}]{Sch:RMP05}
\bibinfo{author}{\bibfnamefont{U.}~\bibnamefont{Schollw{\"o}ck}},
  \bibinfo{journal}{Rev. Mod. Phys.} \textbf{\bibinfo{volume}{77}},
  \bibinfo{pages}{259} (\bibinfo{year}{2005}).

\bibitem[{\citenamefont{White}(1992)}]{Whi:PRL92}
\bibinfo{author}{\bibfnamefont{S.~R.} \bibnamefont{White}},
  \bibinfo{journal}{Phys. Rev. Lett.} \textbf{\bibinfo{volume}{69}},
  \bibinfo{pages}{2863} (\bibinfo{year}{1992}).

\bibitem[{\citenamefont{Ramasesha et~al.}(1997)\citenamefont{Ramasesha, Pati,
  Krishnamurthy, Shuai, and Br\'edas}}]{CorrVecRamasesha}
\bibinfo{author}{\bibfnamefont{S.}~\bibnamefont{Ramasesha}},
  \bibinfo{author}{\bibfnamefont{S.~K.} \bibnamefont{Pati}},
  \bibinfo{author}{\bibfnamefont{H.~R.} \bibnamefont{Krishnamurthy}},
  \bibinfo{author}{\bibfnamefont{Z.}~\bibnamefont{Shuai}}, \bibnamefont{and}
  \bibinfo{author}{\bibfnamefont{J.~L.} \bibnamefont{Br\'edas}},
  \bibinfo{journal}{Synth. Met.} \textbf{\bibinfo{volume}{85}},
  \bibinfo{pages}{1019 } (\bibinfo{year}{1997}), ISSN
  \bibinfo{issn}{0379-6779}.

\bibitem[{\citenamefont{K\"uhner and White}(1999)}]{CorrVecKuehnerWhite}
\bibinfo{author}{\bibfnamefont{T.~D.} \bibnamefont{K\"uhner}} \bibnamefont{and}
  \bibinfo{author}{\bibfnamefont{S.~R.} \bibnamefont{White}},
  \bibinfo{journal}{Phys. Rev. B} \textbf{\bibinfo{volume}{60}},
  \bibinfo{pages}{335} (\bibinfo{year}{1999}).

\bibitem[{\citenamefont{Jeckelmann}(2002)}]{DDMRG}
\bibinfo{author}{\bibfnamefont{E.}~\bibnamefont{Jeckelmann}},
  \bibinfo{journal}{Phys. Rev. B} \textbf{\bibinfo{volume}{66}},
  \bibinfo{pages}{045114} (\bibinfo{year}{2002}).

\bibitem[{\citenamefont{Waldmann et~al.}(1999)\citenamefont{Waldmann,
  Sch\"ulein, Koch, M\"uller, Bernt, Saalfrank, Andres, G\"udel, and
  Allenspach}}]{Wal99-fe6}
\bibinfo{author}{\bibfnamefont{O.}~\bibnamefont{Waldmann}},
  \bibinfo{author}{\bibfnamefont{J.}~\bibnamefont{Sch\"ulein}},
  \bibinfo{author}{\bibfnamefont{R.}~\bibnamefont{Koch}},
  \bibinfo{author}{\bibfnamefont{P.}~\bibnamefont{M\"uller}},
  \bibinfo{author}{\bibfnamefont{I.}~\bibnamefont{Bernt}},
  \bibinfo{author}{\bibfnamefont{R.~W.} \bibnamefont{Saalfrank}},
  \bibinfo{author}{\bibfnamefont{H.~P.} \bibnamefont{Andres}},
  \bibinfo{author}{\bibfnamefont{H.~U.} \bibnamefont{G\"udel}},
  \bibnamefont{and}
  \bibinfo{author}{\bibfnamefont{P.}~\bibnamefont{Allenspach}},
  \bibinfo{journal}{Inorg. Chem.} \textbf{\bibinfo{volume}{38}},
  \bibinfo{pages}{5879} (\bibinfo{year}{1999}).

\bibitem[{\citenamefont{Ochsenbein et~al.}(2008)\citenamefont{Ochsenbein, Tuna,
  Rancan, Davies, Muryn, Waldmann, Bircher, Sieber, Carver, Mutka
  et~al.}}]{Och08-cr6cr7}
\bibinfo{author}{\bibfnamefont{S.~T.} \bibnamefont{Ochsenbein}},
  \bibinfo{author}{\bibfnamefont{F.}~\bibnamefont{Tuna}},
  \bibinfo{author}{\bibfnamefont{M.}~\bibnamefont{Rancan}},
  \bibinfo{author}{\bibfnamefont{R.~S.~G.} \bibnamefont{Davies}},
  \bibinfo{author}{\bibfnamefont{C.~A.} \bibnamefont{Muryn}},
  \bibinfo{author}{\bibfnamefont{O.}~\bibnamefont{Waldmann}},
  \bibinfo{author}{\bibfnamefont{R.}~\bibnamefont{Bircher}},
  \bibinfo{author}{\bibfnamefont{A.}~\bibnamefont{Sieber}},
  \bibinfo{author}{\bibfnamefont{G.}~\bibnamefont{Carver}},
  \bibinfo{author}{\bibfnamefont{H.}~\bibnamefont{Mutka}},
  \bibnamefont{et~al.}, \bibinfo{journal}{Chem. Eur. J.}
  \textbf{\bibinfo{volume}{14}}, \bibinfo{pages}{5144} (\bibinfo{year}{2008}).

\bibitem[{\citenamefont{Stuiber et~al.}(2011)\citenamefont{Stuiber, Wu,
  Nehrkorn, Dreiser, Lan, Novitchi, Anson, Unruh, Powell, and
  Waldmann}}]{Stu11-mn10}
\bibinfo{author}{\bibfnamefont{S.}~\bibnamefont{Stuiber}},
  \bibinfo{author}{\bibfnamefont{G.}~\bibnamefont{Wu}},
  \bibinfo{author}{\bibfnamefont{J.}~\bibnamefont{Nehrkorn}},
  \bibinfo{author}{\bibfnamefont{J.}~\bibnamefont{Dreiser}},
  \bibinfo{author}{\bibfnamefont{Y.}~\bibnamefont{Lan}},
  \bibinfo{author}{\bibfnamefont{G.}~\bibnamefont{Novitchi}},
  \bibinfo{author}{\bibfnamefont{C.~E.} \bibnamefont{Anson}},
  \bibinfo{author}{\bibfnamefont{T.}~\bibnamefont{Unruh}},
  \bibinfo{author}{\bibfnamefont{A.~K.} \bibnamefont{Powell}},
  \bibnamefont{and} \bibinfo{author}{\bibfnamefont{O.}~\bibnamefont{Waldmann}},
  \bibinfo{journal}{Chem. Eur. J.} \textbf{\bibinfo{volume}{17}},
  \bibinfo{pages}{9094} (\bibinfo{year}{2011}).

\bibitem[{\citenamefont{Waldmann}(2003)}]{Wal03-insqdependence}
\bibinfo{author}{\bibfnamefont{O.}~\bibnamefont{Waldmann}},
  \bibinfo{journal}{Phys. Rev. B} \textbf{\bibinfo{volume}{68}},
  \bibinfo{pages}{174406} (\bibinfo{year}{2003}).

\bibitem[{\citenamefont{Furrer et~al.}(2009)\citenamefont{Furrer, Mesot, and
  Str\"assle}}]{ts_insformel}
\bibinfo{author}{\bibfnamefont{A.}~\bibnamefont{Furrer}},
  \bibinfo{author}{\bibfnamefont{J.}~\bibnamefont{Mesot}}, \bibnamefont{and}
  \bibinfo{author}{\bibfnamefont{T.}~\bibnamefont{Str\"assle}},
  \emph{\bibinfo{title}{Neutron Scattering in Condensed Matter Physics}}
  (\bibinfo{publisher}{World Scientific Publishing Co.}, \bibinfo{year}{2009}).

\bibitem[{\citenamefont{Carretta et~al.}(2003)\citenamefont{Carretta, van
  Slageren, Guidi, Liviotti, Mondelli, Rovai, Cornia, Dearden, Affronte, Frost
  et~al.}}]{Car03-cr8}
\bibinfo{author}{\bibfnamefont{S.}~\bibnamefont{Carretta}},
  \bibinfo{author}{\bibfnamefont{J.}~\bibnamefont{van Slageren}},
  \bibinfo{author}{\bibfnamefont{T.}~\bibnamefont{Guidi}},
  \bibinfo{author}{\bibfnamefont{E.}~\bibnamefont{Liviotti}},
  \bibinfo{author}{\bibfnamefont{C.}~\bibnamefont{Mondelli}},
  \bibinfo{author}{\bibfnamefont{D.}~\bibnamefont{Rovai}},
  \bibinfo{author}{\bibfnamefont{A.}~\bibnamefont{Cornia}},
  \bibinfo{author}{\bibfnamefont{C.~F.} \bibnamefont{Dearden},
  \bibfnamefont{A.~L.}},
  \bibinfo{author}{\bibfnamefont{M.}~\bibnamefont{Affronte}},
  \bibinfo{author}{\bibfnamefont{C.~D.} \bibnamefont{Frost}},
  \bibnamefont{et~al.}, \bibinfo{journal}{Phys. Rev. B}
  \textbf{\bibinfo{volume}{67}}, \bibinfo{pages}{094405}
  (\bibinfo{year}{2003}).

\bibitem[{\citenamefont{Waldmann et~al.}(2005)\citenamefont{Waldmann, Dobe,
  Mutka, Furrer, and G\"udel}}]{Wal05-csfe8ins1}
\bibinfo{author}{\bibfnamefont{O.}~\bibnamefont{Waldmann}},
  \bibinfo{author}{\bibfnamefont{C.}~\bibnamefont{Dobe}},
  \bibinfo{author}{\bibfnamefont{H.}~\bibnamefont{Mutka}},
  \bibinfo{author}{\bibfnamefont{A.}~\bibnamefont{Furrer}}, \bibnamefont{and}
  \bibinfo{author}{\bibfnamefont{H.~U.} \bibnamefont{G\"udel}},
  \bibinfo{journal}{Phys. Rev. Lett.} \textbf{\bibinfo{volume}{95}},
  \bibinfo{pages}{057202} (\bibinfo{year}{2005}).

\bibitem[{\citenamefont{Santini et~al.}(2005)\citenamefont{Santini, Carretta,
  Amoretti, Guidi, Caciuffo, Caneschi, Rovai, Qiu, and Copley}}]{San05-fe10nvt}
\bibinfo{author}{\bibfnamefont{P.}~\bibnamefont{Santini}},
  \bibinfo{author}{\bibfnamefont{S.}~\bibnamefont{Carretta}},
  \bibinfo{author}{\bibfnamefont{G.}~\bibnamefont{Amoretti}},
  \bibinfo{author}{\bibfnamefont{T.}~\bibnamefont{Guidi}},
  \bibinfo{author}{\bibfnamefont{R.}~\bibnamefont{Caciuffo}},
  \bibinfo{author}{\bibfnamefont{A.}~\bibnamefont{Caneschi}},
  \bibinfo{author}{\bibfnamefont{D.}~\bibnamefont{Rovai}},
  \bibinfo{author}{\bibfnamefont{Y.}~\bibnamefont{Qiu}}, \bibnamefont{and}
  \bibinfo{author}{\bibfnamefont{J.~R.~D.} \bibnamefont{Copley}},
  \bibinfo{journal}{Phys. Rev. B} \textbf{\bibinfo{volume}{71}},
  \bibinfo{pages}{184405} (\bibinfo{year}{2005}).

\bibitem[{\citenamefont{Waldmann et~al.}(2006)\citenamefont{Waldmann, Dobe,
  G\"udel, and Mutka}}]{Wal06-csfe8ins2}
\bibinfo{author}{\bibfnamefont{O.}~\bibnamefont{Waldmann}},
  \bibinfo{author}{\bibfnamefont{C.}~\bibnamefont{Dobe}},
  \bibinfo{author}{\bibfnamefont{H.~U.} \bibnamefont{G\"udel}},
  \bibnamefont{and} \bibinfo{author}{\bibfnamefont{H.}~\bibnamefont{Mutka}},
  \bibinfo{journal}{Phys. Rev. B} \textbf{\bibinfo{volume}{74}},
  \bibinfo{pages}{054429} (\bibinfo{year}{2006}).

\bibitem[{\citenamefont{Schmidt et~al.}(2001)\citenamefont{Schmidt, Schnack,
  and Luban}}]{SSL:PRB01}
\bibinfo{author}{\bibfnamefont{H.-J.} \bibnamefont{Schmidt}},
  \bibinfo{author}{\bibfnamefont{J.}~\bibnamefont{Schnack}}, \bibnamefont{and}
  \bibinfo{author}{\bibfnamefont{M.}~\bibnamefont{Luban}},
  \bibinfo{journal}{Phys. Rev. B} \textbf{\bibinfo{volume}{64}},
  \bibinfo{pages}{224415} (\bibinfo{year}{2001}).

\bibitem[{\citenamefont{Ivanov and Sen}(2004)}]{IvS:LNP04}
\bibinfo{author}{\bibfnamefont{N.~B.} \bibnamefont{Ivanov}} \bibnamefont{and}
  \bibinfo{author}{\bibfnamefont{D.}~\bibnamefont{Sen}},
  \emph{\bibinfo{title}{Spin Wave Analysis of Heisenberg Magnets in Restricted
  Geometries}} (\bibinfo{publisher}{Springer}, \bibinfo{address}{Berlin,
  Heidelberg}, \bibinfo{year}{2004}), vol. \bibinfo{volume}{645} of
  \emph{\bibinfo{series}{Lecture Notes in Physics}}, chap.
  \bibinfo{chapter}{Molecular Magnetism}, pp. \bibinfo{pages}{195--226}.

\bibitem[{\citenamefont{Anderson}(1952)}]{And52-swt}
\bibinfo{author}{\bibfnamefont{P.~W.} \bibnamefont{Anderson}},
  \bibinfo{journal}{Phys. Rev.} \textbf{\bibinfo{volume}{86}},
  \bibinfo{pages}{694} (\bibinfo{year}{1952}).

\bibitem[{\citenamefont{Oguchi}(1960)}]{oguchi}
\bibinfo{author}{\bibfnamefont{T.}~\bibnamefont{Oguchi}},
  \bibinfo{journal}{Phys. Rev.} \textbf{\bibinfo{volume}{117}},
  \bibinfo{pages}{117} (\bibinfo{year}{1960}).

\bibitem[{\citenamefont{Takahashi}(1989)}]{sw1}
\bibinfo{author}{\bibfnamefont{M.}~\bibnamefont{Takahashi}},
  \bibinfo{journal}{Phys. Rev. B} \textbf{\bibinfo{volume}{40}},
  \bibinfo{pages}{2494} (\bibinfo{year}{1989}).

\bibitem[{\citenamefont{Hirsch and Tang}(1989)}]{HT:PRB1989}
\bibinfo{author}{\bibfnamefont{J.~E.} \bibnamefont{Hirsch}} \bibnamefont{and}
  \bibinfo{author}{\bibfnamefont{S.}~\bibnamefont{Tang}},
  \bibinfo{journal}{Phys. Rev. B} \textbf{\bibinfo{volume}{40}},
  \bibinfo{pages}{4769} (\bibinfo{year}{1989}).

\bibitem[{\citenamefont{Tang et~al.}(1989)\citenamefont{Tang, Lazzouni, and
  Hirsch}}]{sw3}
\bibinfo{author}{\bibfnamefont{S.}~\bibnamefont{Tang}},
  \bibinfo{author}{\bibfnamefont{M.~E.} \bibnamefont{Lazzouni}},
  \bibnamefont{and} \bibinfo{author}{\bibfnamefont{J.~E.}
  \bibnamefont{Hirsch}}, \bibinfo{journal}{Phys. Rev. B}
  \textbf{\bibinfo{volume}{40}}, \bibinfo{pages}{5000} (\bibinfo{year}{1989}).

\bibitem[{\citenamefont{Yamamoto and Hori}(2003)}]{imswt}
\bibinfo{author}{\bibfnamefont{S.}~\bibnamefont{Yamamoto}} \bibnamefont{and}
  \bibinfo{author}{\bibfnamefont{H.}~\bibnamefont{Hori}},
  \bibinfo{journal}{Journal of the Physical Society of Japan}
  \textbf{\bibinfo{volume}{72}}, \bibinfo{pages}{769} (\bibinfo{year}{2003}).

\bibitem[{\citenamefont{Auerbach and Arovas}(1988)}]{SBMFT1}
\bibinfo{author}{\bibfnamefont{A.}~\bibnamefont{Auerbach}} \bibnamefont{and}
  \bibinfo{author}{\bibfnamefont{D.~P.} \bibnamefont{Arovas}},
  \bibinfo{journal}{Phys. Rev. Lett.} \textbf{\bibinfo{volume}{61}},
  \bibinfo{pages}{617} (\bibinfo{year}{1988}).

\bibitem[{\citenamefont{Sarker et~al.}(1989)\citenamefont{Sarker, Jayaprakash,
  Krishnamurthy, and Ma}}]{SBMFT2}
\bibinfo{author}{\bibfnamefont{S.}~\bibnamefont{Sarker}},
  \bibinfo{author}{\bibfnamefont{C.}~\bibnamefont{Jayaprakash}},
  \bibinfo{author}{\bibfnamefont{H.~R.} \bibnamefont{Krishnamurthy}},
  \bibnamefont{and} \bibinfo{author}{\bibfnamefont{M.}~\bibnamefont{Ma}},
  \bibinfo{journal}{Phys. Rev. B} \textbf{\bibinfo{volume}{40}},
  \bibinfo{pages}{5028} (\bibinfo{year}{1989}).

\bibitem[{\citenamefont{Gu et~al.}(2006)\citenamefont{Gu, Su, and
  Gao}}]{GSG:PRB2006}
\bibinfo{author}{\bibfnamefont{B.}~\bibnamefont{Gu}},
  \bibinfo{author}{\bibfnamefont{G.}~\bibnamefont{Su}}, \bibnamefont{and}
  \bibinfo{author}{\bibfnamefont{S.}~\bibnamefont{Gao}},
  \bibinfo{journal}{Phys. Rev. B} \textbf{\bibinfo{volume}{73}},
  \bibinfo{pages}{134427} (\bibinfo{year}{2006}).

\bibitem[{\citenamefont{Waldmann}(2002{\natexlab{b}})}]{Wal02-wheel-qt}
\bibinfo{author}{\bibfnamefont{O.}~\bibnamefont{Waldmann}},
  \bibinfo{journal}{Europhys. Lett.} \textbf{\bibinfo{volume}{60}},
  \bibinfo{pages}{302} (\bibinfo{year}{2002}{\natexlab{b}}).

\bibitem[{\citenamefont{Engelhardt and Luban}(2006)}]{EnL:PRB06}
\bibinfo{author}{\bibfnamefont{L.}~\bibnamefont{Engelhardt}} \bibnamefont{and}
  \bibinfo{author}{\bibfnamefont{M.}~\bibnamefont{Luban}},
  \bibinfo{journal}{Phys. Rev. B} \textbf{\bibinfo{volume}{73}},
  \bibinfo{pages}{054430} (\bibinfo{year}{2006}).

\bibitem[{\citenamefont{Troyer}(1999)}]{Tro:LNCS1999}
\bibinfo{author}{\bibfnamefont{M.}~\bibnamefont{Troyer}},
  \bibinfo{journal}{Lecture Notes in Computer Science}
  \textbf{\bibinfo{volume}{1732}}, \bibinfo{pages}{164} (\bibinfo{year}{1999}).

\bibitem[{\citenamefont{Waldmann et~al.}(2001)\citenamefont{Waldmann, Koch,
  Schromm, Sch\"ulein, M\"uller, Bernt, Saalfrank, Hampel, and
  Balthes}}]{Wal01-csfe8torque}
\bibinfo{author}{\bibfnamefont{O.}~\bibnamefont{Waldmann}},
  \bibinfo{author}{\bibfnamefont{R.}~\bibnamefont{Koch}},
  \bibinfo{author}{\bibfnamefont{S.}~\bibnamefont{Schromm}},
  \bibinfo{author}{\bibfnamefont{J.}~\bibnamefont{Sch\"ulein}},
  \bibinfo{author}{\bibfnamefont{P.}~\bibnamefont{M\"uller}},
  \bibinfo{author}{\bibfnamefont{I.}~\bibnamefont{Bernt}},
  \bibinfo{author}{\bibfnamefont{R.~W.} \bibnamefont{Saalfrank}},
  \bibinfo{author}{\bibfnamefont{F.}~\bibnamefont{Hampel}}, \bibnamefont{and}
  \bibinfo{author}{\bibfnamefont{E.}~\bibnamefont{Balthes}},
  \bibinfo{journal}{Inorg. Chem.} \textbf{\bibinfo{volume}{40}},
  \bibinfo{pages}{2986} (\bibinfo{year}{2001}).

\bibitem[{\citenamefont{Takahashi}(1987)}]{takahashi}
\bibinfo{author}{\bibfnamefont{M.}~\bibnamefont{Takahashi}},
  \bibinfo{journal}{Phys. Rev. Lett.} \textbf{\bibinfo{volume}{58}},
  \bibinfo{pages}{168} (\bibinfo{year}{1987}).

\bibitem[{\citenamefont{M\"uller}(1982)}]{Mue:PRB1982}
\bibinfo{author}{\bibfnamefont{G.}~\bibnamefont{M\"uller}},
  \bibinfo{journal}{Phys. Rev. B} \textbf{\bibinfo{volume}{26}},
  \bibinfo{pages}{1311} (\bibinfo{year}{1982}).

\bibitem[{\citenamefont{Waldmann and G\"udel}(2005)}]{Wal05-smix}
\bibinfo{author}{\bibfnamefont{O.}~\bibnamefont{Waldmann}} \bibnamefont{and}
  \bibinfo{author}{\bibfnamefont{H.~U.} \bibnamefont{G\"udel}},
  \bibinfo{journal}{Phys. Rev. B} \textbf{\bibinfo{volume}{72}},
  \bibinfo{pages}{094422} (\bibinfo{year}{2005}).

\bibitem[{\citenamefont{Nishimoto and Arikawa}(2009)}]{DDMRGNishimoto}
\bibinfo{author}{\bibfnamefont{S.}~\bibnamefont{Nishimoto}} \bibnamefont{and}
  \bibinfo{author}{\bibfnamefont{M.}~\bibnamefont{Arikawa}},
  \bibinfo{journal}{Phys. Rev. B} \textbf{\bibinfo{volume}{79}},
  \bibinfo{pages}{113106} (\bibinfo{year}{2009}).

\bibitem[{\citenamefont{Honecker et~al.}(2011)\citenamefont{Honecker, Hu,
  Peters, and Richter}}]{DDMRGPeters}
\bibinfo{author}{\bibfnamefont{A.}~\bibnamefont{Honecker}},
  \bibinfo{author}{\bibfnamefont{S.}~\bibnamefont{Hu}},
  \bibinfo{author}{\bibfnamefont{R.}~\bibnamefont{Peters}}, \bibnamefont{and}
  \bibinfo{author}{\bibfnamefont{J.}~\bibnamefont{Richter}},
  \bibinfo{journal}{J. Phys.: Condens. Matter} \textbf{\bibinfo{volume}{23}},
  \bibinfo{pages}{164211} (\bibinfo{year}{2011}).

\bibitem[{\citenamefont{Exler and Schnack}(2003)}]{ExS:PRB03}
\bibinfo{author}{\bibfnamefont{M.}~\bibnamefont{Exler}} \bibnamefont{and}
  \bibinfo{author}{\bibfnamefont{J.}~\bibnamefont{Schnack}},
  \bibinfo{journal}{Phys. Rev. B} \textbf{\bibinfo{volume}{67}},
  \bibinfo{pages}{094440} (\bibinfo{year}{2003}).

\bibitem[{\citenamefont{Yang}(2000)}]{Yan:Book2000}
\bibinfo{author}{\bibfnamefont{D.}~\bibnamefont{Yang}},
  \emph{\bibinfo{title}{{C++ and Object-oriented Numeric Computing for
  Scientists and Engineers}}} (\bibinfo{publisher}{Springer Verlag},
  \bibinfo{address}{New York Berlin Heidelberg}, \bibinfo{year}{2000}).

\bibitem[{\citenamefont{Schneider et~al.}(2008)\citenamefont{Schneider, Struck,
  Bortz, and Eggert}}]{SSBE:PRL2008}
\bibinfo{author}{\bibfnamefont{I.}~\bibnamefont{Schneider}},
  \bibinfo{author}{\bibfnamefont{A.}~\bibnamefont{Struck}},
  \bibinfo{author}{\bibfnamefont{M.}~\bibnamefont{Bortz}}, \bibnamefont{and}
  \bibinfo{author}{\bibfnamefont{S.}~\bibnamefont{Eggert}},
  \bibinfo{journal}{Phys. Rev. Lett.} \textbf{\bibinfo{volume}{101}},
  \bibinfo{pages}{206401} (\bibinfo{year}{2008}).

\bibitem[{\citenamefont{Lieb et~al.}(1961)\citenamefont{Lieb, Schultz, and
  Mattis}}]{LSM:AP61}
\bibinfo{author}{\bibfnamefont{E.~H.} \bibnamefont{Lieb}},
  \bibinfo{author}{\bibfnamefont{T.}~\bibnamefont{Schultz}}, \bibnamefont{and}
  \bibinfo{author}{\bibfnamefont{D.~C.} \bibnamefont{Mattis}},
  \bibinfo{journal}{Ann. Phys. (N.Y.)} \textbf{\bibinfo{volume}{16}},
  \bibinfo{pages}{407} (\bibinfo{year}{1961}).

\bibitem[{\citenamefont{Lieb and Mattis}(1962)}]{LiM:JMP62}
\bibinfo{author}{\bibfnamefont{E.~H.} \bibnamefont{Lieb}} \bibnamefont{and}
  \bibinfo{author}{\bibfnamefont{D.~C.} \bibnamefont{Mattis}},
  \bibinfo{journal}{J.~Math. Phys.} \textbf{\bibinfo{volume}{3}},
  \bibinfo{pages}{749} (\bibinfo{year}{1962}).

\end{thebibliography}

\end{document}